\documentclass[prl, aps, superscriptaddress, longbibliography,showpacs,amsmath,amssymb,floatfix, 10pt, twocolumn]{revtex4-1}

\usepackage{comment}
\usepackage{graphicx}
\usepackage{dcolumn}
\usepackage{bm}
\usepackage{graphicx}             
\usepackage{booktabs}                                  
\usepackage{amssymb}

\usepackage{amsmath}
\usepackage{epsfig}
\usepackage{xcolor}
\usepackage{tabu}
\usepackage{mathtools}
\usepackage[colorlinks,linkcolor=blue,anchorcolor=blue,citecolor=blue,urlcolor=blue]{hyperref}
\usepackage{physics}
\usepackage{float}
\usepackage{diagbox}
\usepackage{inputenc}

\begin{document}

\title{Cold atomic ensembles as quantum antennas for distributed networks of single-atom arrays}
\author{Xiaoshui Lin}
\affiliation{Department of Physics, Washington University, St. Louis, MO, 63130, USA}

\author{Yefeng Mei}
\affiliation{Department of Physics and Astronomy, Washington State University, Pullman, Washington 99164, USA}

\author{Chuanwei Zhang}
\email{chuanwei.zhang@wustl.edu}
\affiliation{Department of Physics, Washington University, St. Louis, MO, 63130, USA}
\affiliation{Department of Physics, University of Texas at Dallas, Richardson, TX, 75080, USA}

\begin{abstract}
Single neutral atoms in optical tweezer arrays offer a promising platform for high-fidelity quantum computing at local nodes.
Nonetheless, creating entanglement between remote nodes in a distributed quantum network remains challenging due to inherently weak atom-light coupling. Here, we design a distributed quantum network architecture in which cold atomic ensembles with strong atom-light interactions act as quantum antennas, interfacing single-atom qubits with flying photons to enable high-efficiency atom-photon entanglement generation---analogous to the role of antennas in classical communication.
Using realistic experimental parameters, we estimate an efficiency of $\eta \simeq 0.548$ for generating atom-photon entanglement, a probability of $P_{E} \simeq  6 \%$ for generating atom-atom entanglement, and a remote entanglement generation rate of $16.6 $ kHz. 
This performance not only surpasses that of state-of-the-art cavity-based or high-numerical-aperture-lens-based architectures but also offers notable advantages in simplicity, tunability, and experimental accessibility. 
Our scheme also integrates a long-lived quantum memory, providing a storage advantage for quantum repeater design.
By leveraging the complementary strengths of single-atom qubits for local operations and cold atomic ensembles for networking, this approach paves the way for scalable distributed quantum computing and sensing.
\end{abstract}

\maketitle 

\begin{figure*}[htbp]
\centering
\includegraphics[width=0.95\linewidth]{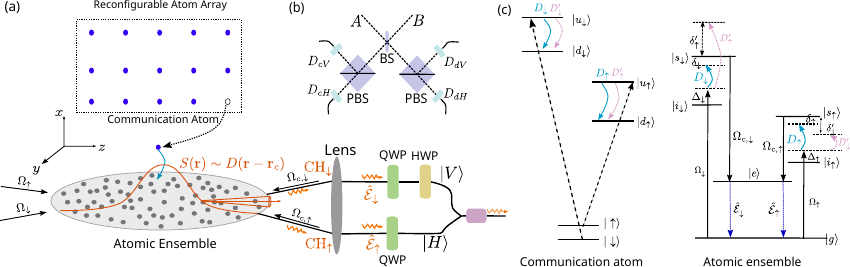}  
\caption{ 
(a) Two types of qubits within a single node. 
Atoms in optical tweezer arrays serve as memory or computation qubits, while a communication atom, together with a cold atomic ensemble, forms the quantum antenna responsible for generating atom-photon entanglement.
The communication atom is spatially separated from the ensemble yet  retains strong dipole-dipole interactions with the ensemble's Rydberg states. 
The photon qubit is encoded in the polarization channel using a dual-rail scheme.
(b) For two distant quantum nodes A and B, the photons are collected at a middle station, pass through a Beam splitter (BS) and then through a polarization Beam splitter (PBS). 
If photons are detected in $D_{cH}$, $D_{cV}$ or $D_{dH}$ and $D_{dV}$, the state of nodes A and B is projected onto $|\Psi^{+} \rangle = (|\downarrow \uparrow \rangle_{AB} + |\uparrow \downarrow \rangle_{AB} )/\sqrt{2}$. 
While the photon detection in $D_{cH}$, $D_{dV}$ or $D_{cV}$ and $D_{dH}$ gives rise to $|\Psi^{-} \rangle = (|\downarrow \uparrow \rangle_{AB} - |\uparrow \downarrow \rangle_{AB} )/\sqrt{2}$.
The success probability of this process is $\frac{1}{2}$, and entanglement is established only when two photons with different polarizations are collected.
(c) Energy level structure of the communication atoms and the atomic ensemble. 
Two hyperfine states of the atom, denoted $| \uparrow \rangle$ and $|\downarrow \rangle$, are used as communication qubits. 
The transition $|u_{\mu} \rangle \rightarrow | d_{\mu}\rangle$ is coupled to the atomic ensemble's Rydberg states $| i_{\mu} \rangle \rightarrow |s_{\mu}\rangle$, through a dipole-dipole exchange interaction $D_{\mu}(\mathbf{r} - \mathbf{r}_s)$ with $\mu \in \{\uparrow, \downarrow\}$. 
The interactions from the paired state $|u_{\mu} \rangle | i_{\mu}\rangle$ to paired state $|d_{\mu} \rangle | s_{\nu}\rangle$ is denoted $D_{\mu}'$, where $\nu = \{\uparrow, \downarrow\}$ for $\mu = \{ \downarrow, \uparrow\}$. 
These effects can be neglected due to the energy offset. 
Laser pulse $\Omega_{\mu}$ with different wave vectors $\mathbf{k}_{\mu,0}$ couple the ground state $|g \rangle$ to the intermediate state $|i_{\mu} \rangle$ with Rabi frequency $\Omega_{\mu}$ and detuning $\Delta_{\mu} \gg \Omega_{\mu}$.
Together with the dipole-dipole exchange $|u_{\mu} \rangle |i_{\mu} \rangle$ to $|d_{\mu}\rangle |s_{\mu} \rangle$, this process leads to a transition $|g \rangle \rightarrow |s_{\mu} \rangle$ detuned by $\protect\delta$.
The resulting spin wave state has a spatial distribution $S_{\mu}(\mathbf{r}) \propto \sqrt{ \rho(\mathbf{r})}D_{\mu}(\mathbf{r}-\mathbf{r}_s)e^{i \mathbf{k}_{\mu,0} \cdot \mathbf{r}} $. 
For converting the spin-wave to a propagating photon, the spin-wave states $|s_{\mu} \rangle$ are resonantly coupled to a low-lying state $|e\rangle$.
The low-lying state $|e\rangle$ decays to the ground state, emitting a photon along the direction $\mathbf{k}_{\mu} = \mathbf{k}_{\mu,0} - \mathbf{k}_{\mu,c}$, as dictated by the phase-matching condition.}
\label{fig1}
\end{figure*}

Neutral atoms trapped in optical tweezer arrays have emerged as a leading platform for scalable quantum computing due to their flexible configurability and full connectivity \cite{Barredo2016Atom, Evered2023High, Bluvstein2024Logical}. As qubits, neutral atoms offer several attractive features \cite{Wang2016SingleQubit, Xia2015Randomized, Weiss2017Quantum, Madjarov2020High}, including long coherence times, high-precision control, and excellent scalability. 
To realize large-scale quantum computation, each quantum node—comprising hundreds of qubits—must be interconnected and entangled with others \cite{Duan2010Colloquium, Sangouard2011Quantum, Wehner2018Quantum, Azuma2023Quantum}. While single neutral atoms are well suited for local operations \cite{Levine2019Parallel}, they pose significant challenges for remote entanglement generation because of their inherently weak atom-light interactions. 
To address this limitation, high-finesse cavities \cite{Duan2004Scalable,Ritter2012Elementary, Reiserer2015Cavity-based, Young2022Architecture, Covey2023Quantum, Reiserer2022Colloquium} and high-numerical-aperture (NA) lenses \cite{Robens2017High, VanLeent2020Long-Distance, Stephenson2020High-Rate} have been employed to enhance photon collection efficiency.
However, it remains a significant challenge to coherently and independently manipulate multiple neutral atoms in optical tweezer arrays inside high-finesse cavities or high-NA lenses in vacuum.

Alternatively, cold atomic ensembles have played a major role in the development of quantum networks, owing to their collectively enhanced atom-light interaction strength and the highly directed photon scattering enabled by phase matching \cite{Duan2001Longdistance, Lukin2003ColloquiumTrappingManipulating, Hammerer2010QuantumInterface, Sangouard2011QuantumRepeatersBased, Sangouard2011Quantum}. However, cold atomic ensembles are not ideal candidates for local quantum nodes due to their short coherence times and low-fidelity quantum gates \cite{Ebert2015Coherence, Xu2021Fast}. Moreover, the absence of local two-qubit gates between ensembles limits the generation of multipartite entanglement in practical quantum networks.

The complementary strengths and weaknesses of single-atom qubits and cold atomic ensembles naturally raise an important question: Can remote quantum nodes composed of single-atom arrays be interconnected via cold atomic ensembles to enable large-scale distributed quantum computation? 
In such an architecture, a cold atomic ensemble serves as an interface between single atoms and photons propagating through free space—analogous to an antenna in classical radio communication, which links electric currents to radio waves. As a quantum antenna, the cold atomic ensemble can both transmit and receive photons, making it as a potentially essential component of future distributed quantum networks.

A key requirement for constructing such a quantum antenna is achieving a coherent and controllable interaction between a single atom in an optical tweezer and a cold atomic ensemble. This interaction enables entanglement between the single-atom qubit and photons scattered from the ensemble. 
Such controlled interaction can be achieved via dipole-dipole exchange interactions \cite{Petrosyan2018Deterministic, Ravets2014Coherent, Beguin2013Direct, Barredo2014Demonstration, Barredo2015Coherent} between the single atom and the collective spin-wave excitation of the cold atomic ensemble. This mechanism enables efficient entanglement between a single-atom qubit and a directionally emitted photon from the ensemble. 
Moreover, the cold atomic ensemble not only functions as a quantum antenna but also serves as a long-lived quantum memory, enabling storage of entanglement during heralding and forming a key building block for quantum repeater designs that require buffering and synchronization across distant network links.

\textit{Quantum antenna architecture}: 
In this Letter, we propose an experimentally feasible scheme to realize such a quantum-antenna architecture and compare its efficiency with that of other quantum network schemes. The structure of a single quantum node in a quantum-antenna–enabled distributed quantum network is schematically illustrated in Fig. \ref{fig1}(a).
Each quantum node contains two types of qubits. The communication qubit comprises a single atom—referred to as the communication atom—paired with a cold-atom ensemble that serves as the quantum antenna. 
The atomic ensemble is confined in a one-dimensional optical lattice operated at the magic wavelength for the relevant Rydberg transition, ensuring state-insensitive trapping and thereby preserving the ground–Rydberg coherence for tens of microseconds \cite{Mei2022Trapped,li2022dynamics,lampen2018long}.
The remaining single atoms, trapped in optical tweezers, function as memory or computation qubits, performing quantum gate operations within the node. These quantum nodes are interconnected via Bell-state measurements on flying photons, as shown in Fig. \ref{fig1}(b), with a success probability of 1/2.
This division between communication and memory qubits enables remote entanglement generation across multiple quantum nodes, paving the way for scalable distributed quantum computing.

\textit{(I) Interface between the ensemble and the single atom:} 
For the single atom, the quantum information is stored in two lower hyperfine states, denoted as $|\downarrow \rangle$ and $| \uparrow \rangle$, as shown in Fig. \ref{fig1} (c).
Two Rydberg states, $|u_{\mu} \rangle$ and $|d_{\mu}\rangle$ of the single atom are used to interact with the atomic ensemble.
For convenience, we define $(\mu, \nu) \in \{(\downarrow, \uparrow), (\uparrow, \downarrow) \}$.
For the atomic ensemble, the ground state $|g \rangle$ and two sets of excited Rydberg states $|i_{\mu} \rangle$ and $|s_{\mu}\rangle$ are utilized for the ensemble-communication-atom interface. 

To generate the interface, we employ a coherent $\pi$ pulse or adiabatic passage techniques to transfer the communication atom from state $|\mu \rangle$ to the Rydberg state $|u_{\mu} \rangle$ with nearly $ 100\%$ fidelity \cite{Bergmann1998Coherent,
Kay2010Perfect}.
Two lasers are then used to couple the ensemble's ground state $|g \rangle$ to its intermediate states $|i_{\mu} \rangle$ with detuning $\Delta_{\mu} = \omega_{\mu} - \omega_{ig,\mu}$, Rabi frequency $\Omega_{\mu} $, and wave vector $\mathbf{k}_{\mu,0}$. 
The direction of $\mathbf{k}_{\uparrow,0}$ and $\mathbf{k}_{\downarrow,0}$ are chosen to differ by a small angle, ensuring that each spin wave has a distinct spatial direction.
The paired Rydberg states $|u_{\mu} \rangle| i_{\mu}\rangle$ for the communication exhibit a strong dipole-dipole interaction $D_{\mu,j}$ with the states $|d_{\mu} \rangle |s_{\mu} \rangle $ and an interaction $D_{\mu,j}'$ with the states $|d_{\mu} \rangle |s_{\nu} \rangle $ for the $j$-th atom in the ensemble.
The strength of the first type of interaction is given by $D_{\mu,j} = D_{\mu}(\mathbf{r}_j - \mathbf{r}_c) = \frac{C_{3,\mu}}{2 |\mathbf{r}_j - \mathbf{r}_c|^3}[1 - 3\cos^2(\theta_j)]$, where $C_{3,\mu}$ is the dipole-dipole interaction constant, $\mathbf{r}_j$ is the position of the atom in the ensemble, $\mathbf{r}_c$ is the position of the communication atom, and $\theta_j $ is the angle between $\mathbf{r}_j - \mathbf{r}_c$ and the quantized axis $z$. 
The formula of the second type of interaction is the same, with a slightly different interaction coefficient $C_{3,\mu}'$. 
We assume the single photon detuning $\Delta_{\mu} \simeq \omega_{si,\mu} - \omega_{ud, \mu}$ to be much larger than the other parameters $\Omega_{\mu}$, $D_{\mu,j}$, $D'_{\mu,j}$, and $\delta_{\mu} = \omega_{\mu} + \omega_{ud,\mu} - \omega_{eg,\mu}$, which enables adiabatic elimination of the intermediate state $|i_{\mu} \rangle$. 
We take the two-photon detuning to be $\delta_{\mu} \sim 10$ MHz, which is small.  Meanwhile, the other two-photon detuning $\delta_{\mu}' = \omega_{\nu} + \omega_{ud,\mu} - \omega_{eg,\nu}  \sim 300$ MHz, is large.
Since $D_{\mu,j}' \ll |\delta_{\mu}' \pm \Delta_{\nu}|$, we show that its contribution to the dynamic process can be neglected (see discussion in \cite{SI}).

Although the energy level structure of Hilbert space for the whole system is complicated, the dynamics starting from the initial state $|u_{\mu} \rangle |g \rangle$ are constrained in certain subspace. 
In this subspace, as we assume $\Delta_{\mu}$ and $\delta_{\mu}'$ are much larger than the other parameters, the intermediate states $|u_{\mu}, i_{\mu} \rangle$, $|u_{\mu}, i_{\nu}\rangle$, and $| d_{\mu}, s_{\nu} \rangle$ can be adiabatically eliminated (see Ref. \cite{SI} for details).
The resulting effective Hamiltonian is given by  
\begin{equation}
\tilde{H}_{0,\mu} = - \tilde{\delta}_{\mu} | d_{\mu}, S_{\mu} \rangle \langle  d_{\mu}, S_{\mu} |  + \bar{D}_{\mu} ( | d_{\mu},S_{\mu} \rangle \langle u_{\mu}, G | + \mathrm{H.c.} ).  \label{eq-effective-two-level}
\end{equation}
Here, the difference in the effective two-photon detuning $\tilde{\delta}_{\mu,j} = \delta_{\mu} + \frac{|\Omega_{\downarrow}|^2}{\Delta_{\downarrow}} +\frac{|\Omega_{\uparrow}|^2}{\Delta_{\uparrow}} - \frac{|D_{\mu,j}|^2}{\Delta_{\mu}} = \tilde{\delta}_{\mu} -  \frac{|D_{\mu,j}|^2}{\Delta_{\mu}}$ is neglected. The parameters are defined as $\bar{D}_{\mu} = \sqrt{\sum_j |\tilde{D}_{\mu,j}|^2} $ with $\tilde{D}_{\mu, j} = \frac{\Omega_{\mu} D_{\mu,j}}{\Delta_{\mu}}$. 
The collective spin-wave state is 
$|S_{\mu} \rangle = \bar{D}_{\mu}^{-1} \sum_j \tilde{D}_{\mu,j} e^{i \mathbf{k}_{\mu,0} \cdot \mathbf{r}_j} |g\rangle_1|g\rangle_2 \dots |s_{\mu} \rangle_j \dots |g \rangle_N$, and the collective ground state is $|G \rangle = |g \rangle_1 |g \rangle_2 \dots |g \rangle_N$.
Explicitly, the dynamics for the initial states $|u_{\downarrow} , G \rangle$ and $|u_{\uparrow}, G \rangle$ are independent. 
By setting the effective two-photon detuning to $\tilde{\delta}_{\mu} = \alpha(t - t_0)$, we can adiabatically convert the state $|u_{\mu} \rangle |G \rangle$ to state $|d_{\mu}\rangle |S_{\mu} \rangle$ with high fidelity \cite{Vitanov2001Laser, SI}.
In this process, the collective coupling strength $\bar{D}_{\mu} \propto \sqrt{N}$ and depends on both the position and number of atoms in the ensemble.
This dependence causes fluctuations across different experimental realizations.
However, since the nonadiabatic transition probability is $P_{\mu} = e^{-2\pi \bar{D}_{\mu}^2/\alpha}$,  the adiabatic condition is satisfied whenever the mean value of $\bar{D}_{\mu}^2$ is much larger than the parameter's rate of change.  
After this adiabatic state transfer process, we employ a $\pi$ pulse to convert the communication from  $|d_{\mu}\rangle$ back to  $|\mu\rangle$.
Thus, if the initial state of the communication atom is chosen as $|\Psi\rangle_c = (|\downarrow \rangle + |\uparrow \rangle)/\sqrt{2}$, the resulting entangled state is 
\begin{equation}
|\Psi_0 \rangle = \frac{1}{\sqrt{2}} (| \downarrow \rangle | S_{\downarrow} \rangle +
e^{i \phi_{d}} | \uparrow \rangle | S_{\uparrow} \rangle ),
\label{eq-atom-ensemble-entangle}
\end{equation}
which is the maximally entangled state between the communication atom and the ensemble. 

\begin{figure}[htbp!]
    \centering
    \includegraphics[width=0.99\linewidth]{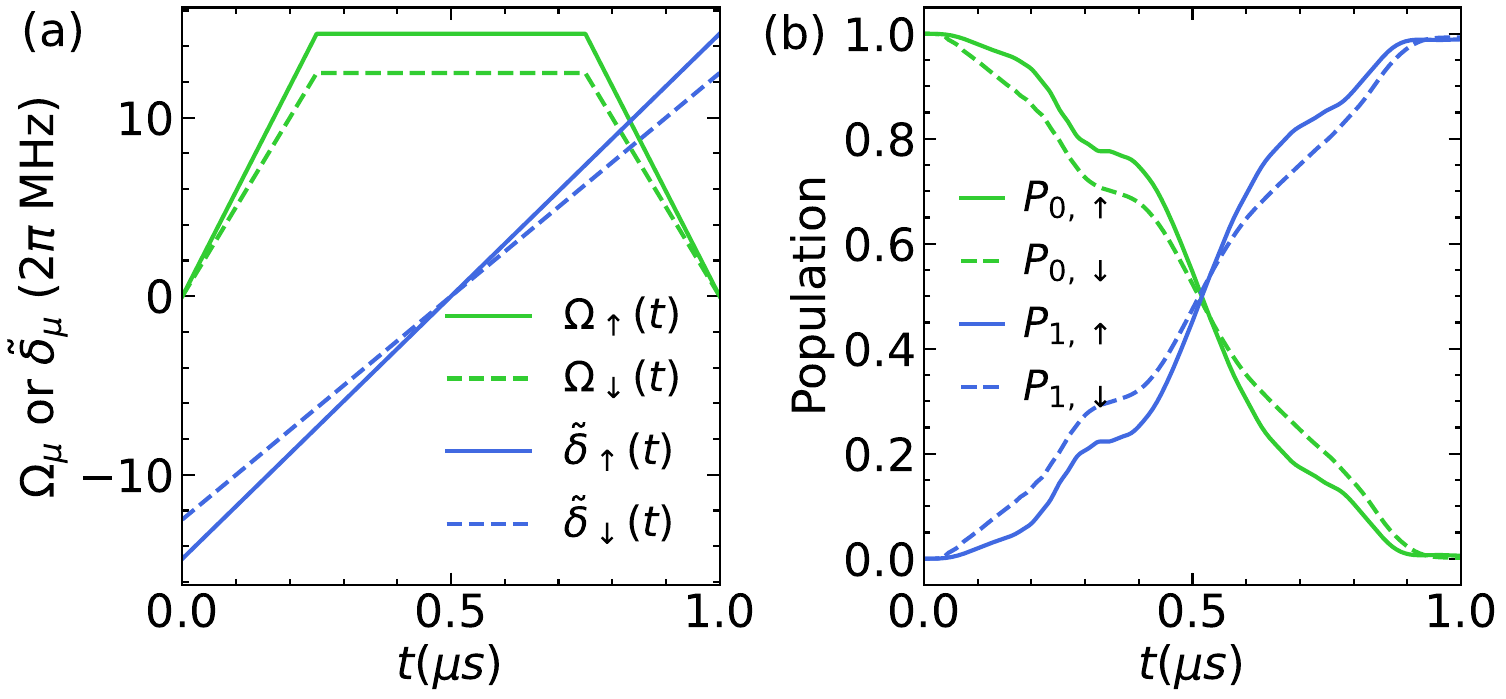}
    \caption{(a) Time dependence of the Rabi frequency $\Omega_{\mu}$ and the effective two-photon detuning $\tilde{\delta}_{\mu}$.
    (b) Dynamics for the population $P_{\mu, 0} = |C_{\mu,0}|^2$ and populations $P_{\mu,1} = \sum_{j=1}|C_{\mu,j}|^2 $ for different initial state $|u_\mu \rangle|G \rangle$. 
    The value for $P_{\mu,1}(t = 4\ \mu s)$ is given by $P_{\uparrow,1}(t = 1\ \mu s) = 0.9893$ and $P_{\downarrow,1}(t = 1\ \mu s) = 0.9931$.
    In the simulation, we use $N = 500$ atoms with a peak density $\rho_\text{peak} \simeq 0.53\ \mu m^{-3}$, $\sigma_{\bot} = 4\ \mu m$, $\sigma_z = 30\ \mu m$, and $x_c \simeq 7.1\ \mu \text{m}$. The decay rate of $|e\rangle$ is set to $\Gamma_s = 2\pi \times 10$ kHz.
    The results are averaged over 100 realizations. 
    For the communication atom, we adopt parameters corresponding to the $^{85}$Rb with $|u_{\mu} \rangle = | m_{\mu} P_{3/2}, F = 4, m_F = -4\rangle$ and $|d_{\mu} \rangle = |m_{\mu} S_{1/2}, F = 4, m_F = -3\rangle$, where $m_{\mu} = 65$ for $\mu = \uparrow$ and $m_{\mu} = 68$ for $\mu = \downarrow$.
    For the atomic ensemble, we adopt parameters corresponding to the $^{85}$Rb with $|i_{\mu} \rangle = | (m_\mu - 1) P_{3/2}, F = 4, m_F = 4\rangle$ and $|s_{\mu} \rangle = | m_{\mu} S_{1/2}, F = 3, m_F = 3\rangle$.
    These choices yield to detunings of $\Delta_{\uparrow} = 2\pi \times 147$ MHz and $\Delta_{\downarrow} = 2\pi \times 125$ MHz. 
    }
    \label{fig2}
\end{figure}

To validate the above analysis, we performed numerical simulations using realistic atomic parameters \cite{SI}.
We place $N \gg 1$ atoms in the ensemble, centered at the origin $(x, y, z) = (0,0,0)$, and Gaussian distributed along all directions with waists $\sigma_x = \sigma_y = \sigma_{\bot}$  and $\sigma_z$. 
The communication atom is placed at $(x_c, 0, 0)$ outside the ensemble cloud (see S5 in \cite{SI}). 
Taking into account the spontaneous decay of the Rydberg state at rate $\Gamma_s$, the resulting population dynamics are shown in Fig. \ref{fig2} (b). We find the writing efficiency for preparing $|S_{\uparrow} \rangle$ is $\eta_{\uparrow, w} \simeq 0.9893$ and that for $|S_{\downarrow} \rangle$ is $\eta_{\downarrow, w} \simeq 0.9931 $ after averaging over 100 realizations. Notably, our scheme enables near-unity single Rydberg excitation in an ensemble without relying on Rydberg blockade, which typically requires highly excited states or strong spatial confinement. (see discussion in \cite{SI}).

\textit{(II) Interface between the photon and the single atom:} 
With the interface between the communication atom and the atomic ensemble established, we can now convert this entanglement into atom–photon entanglement.   
The key idea in this step is to employ the electromagnetically induced transparency (EIT) protocol \cite{Fleischhauer2000Dark-state, Fleischhauer2005Electromagnetically, Finkelstein2023Practical, Novikova2012Electromagnetically, Petrosyan2011Electromagnetically}. 
We resonantly couple the state $|s_{\mu} \rangle$ to a low-lying state $|e\rangle$ using two lasers with Rabi frequency $\Omega_{c,\mu}$ and wave vector $\mathbf{k}_{c,\mu}$. 
Then, the low-lying state $|e\rangle$ is coupled to the ground state $|g \rangle$ via  quantaized electromagnetic field $\hat{\mathcal{E}}(\mathbf{r}, t) = \sum_{k} \hat{a}_{k}(t) \mathrm{e}^{\mathrm{i}(\mathbf{k} - \omega_{eg}/c) \cdot \mathbf{r}} e^{i \nu t}$. 
Furthermore, we set the transition process from $|e\rangle$ to $|g\rangle$ to be mediated only by $\sigma^{+}$ photons.  
To avoid unwanted transition between the spin-wave states $|s_{\uparrow} \rangle$ and $|s_{\downarrow} \rangle$ during retrieval, the control lasers $\Omega_{\downarrow}$ and $\Omega_{\uparrow}$ are switched on according to a designed time sequence, while a phase-locking system actively stabilizes the relative phase between $\Omega_{c,\downarrow}$ and $\Omega_{c,\uparrow}$, as in the dual-rail scheme \cite{Vernaz2018Highly, Wang2019Efficient}. 
For each retrieval process with a single photon excitation, we define a dark-state polarization operator 
\begin{equation}
    \hat{\Psi}_{\mu}(z, t) = \cos(\theta(t)) \hat{\mathcal{E}}(z, t) - \sin(\theta(t)) \hat{S}_{\mu}(z, t),   
\end{equation}
where $ \cos(\theta) = \Omega_c(t)/\sqrt{[\Omega_c(t)]^2 + g^2 N}$, $\quad \sin(\theta) = g \sqrt{N}/\sqrt{[\Omega_c(t)]^2 + g^2 N}$, and $\hat{S}_{\mu}(z,t) = \sqrt{N} \hat{\sigma}_{eg}(z)$.  
The equation for this dark-state polarization operator is
\begin{equation}
    [\partial_t + c \cos^2(\theta(t)) \partial_z] \hat{\Psi}(x, t) = 0. 
\end{equation}
The solution for this polarization operator is generally 
\begin{equation}
    \hat{\Psi}(z, t) = \hat{\Psi}\left(z - \int_{t_0}^{t} dt' c \cos^{2}[\theta(t')] , t = t_0\right).
\end{equation}
Thus, using an appropriate control laser $\Omega_c(t)$, we can convert the spin-wave state $|S_{\mu} \rangle$ of the ensemble into a well-directed single photon propagating along $\mathbf{k}_{\mu,0} - \mathbf{k}_{\mu,c}$, as shown in Fig. \ref{fig1} (a) \cite{Lukin2003ColloquiumTrappingManipulating, Gorshkov2007Photon, Gorshkov2007Universal}.
Thus, starting from the state in Eq. \ref{eq-atom-ensemble-entangle}, the retrieval process yields an entangled state between the communication atom and the propagating photon $|\tilde{\Psi}_0 \rangle = \frac{1}{\sqrt{2}} (|\downarrow \rangle | \text{CH}_{\downarrow} \rangle +  e^{i \phi_{d}'} |\uparrow \rangle | \text{CH}_{\uparrow}   \rangle )$. To further encode the photon information into the polarization channel, we collect the photons via a lens and  pass them through a quarter-wave plate (QWP) to convert them into horizontally linear-polarized photons. 
In particular, for $\text{CH}_{\uparrow}$, an additional half-wave plate (HWP) is used to rotate the polarization to vertical.
Thus, combining all these processes, we obtain the final entangled state between the communication atom and the polarized photon 
\begin{equation}
  |\Psi_1 \rangle = \frac{1}{\sqrt{2}} (|\downarrow \rangle | V \rangle +  e^{i \phi_{d}'} |\uparrow \rangle | H \rangle ).
  \label{eq-atom-photon-entangle}
\end{equation}
Here $\phi_{d}'$ is the dynamic phase. 

In the experiment, the probability of converting the spin-wave excitation into a well-defined propagating photon is limited by the optical depth (OD). For an atomic ensemble with the same parameters as in Fig. \ref{fig2}, we obtain $\text{OD} \simeq 6$ \cite{SI}. 
Following the method in Refs. \cite{Gorshkov2007Universal, Gorshkov2007Photon, Ornelas2020Demand,Mei2022Trapped}, we estimate a retrieval efficiency of $\eta_{r} \simeq 0.55 $ for $\Gamma_{eg} = 6$ MHz and $\Omega_c = 2\pi \times 10$ MHz.
Combining the efficiency of entangling the communication atom with the atomic ensemble, the total efficiency is $\eta = \eta_{r} \eta_{w} \simeq  0.548$. Here, we neglect the slight difference between $\eta_{\mu,w}$ for different $\mu$. 
This efficiency can be further increased by using an atomic ensemble with higher OD. For example, with $\text{OD} = 13$ \cite{Ornelas2020Demand}, the generation efficiency could reach $\eta \simeq 0.69$. 

\textit{(III) Generation of remote entanglement between two distant quantum nodes:}
With an interface between the communication atom and the propagating photon, we can establish remote entanglement through a Bell measurement, similar to those used in quantum teleportation \cite{Pirandola2015Advances, Hu2023Progress}. 
As shown in Fig. \ref{fig1} (b), we first establish the interface between the communication atom and the photon in both nodes A and B.
The photons are then collected and transmitted via optical fiber to a middle station.
If the two photons have the same polarization, they will always exit the beam splitter (BS) together in the same output mode due to the Hong–Ou–Mandel effect \cite{Hong1987Measurement, Chen2015Measuring}.
In this case, one of the four single-photon detectors will register both photons, and no remote entanglement is established.
In contrast, if the two photons have different polarizations, they can exit the beam splitter in either the same or different output modes.
For the same output mode, photon detection projects the communication atoms in the two quantum nodes onto $|\Psi^{+} \rangle = (|\downarrow \uparrow \rangle_{AB} + |\uparrow \downarrow \rangle_{AB} )/\sqrt{2}$ \cite{Duan2001Longdistance, Sangouard2011QuantumRepeatersBased, Yuan2008Experimental, Li2019Experimental}.
For different output modes, the measurement projects the wave function of nodes A and B onto $|\Psi^{+} \rangle = (|\downarrow \uparrow \rangle_{AB} - |\uparrow \downarrow \rangle_{AB} )/\sqrt{2}$. 
Here, we neglect the relative phase between the states $|\downarrow \uparrow \rangle$ and $|\uparrow \downarrow \rangle$, as it does not affect the entanglement properties of the wave function for nodes A and B. 

\begin{table}[t]
    \centering
    \begin{tabular}{ccccc}
    \hline
         & Cavity \cite{Niemietz2021Nondestructive} & Cavity \cite{Young2022Architecture} & HNA lens \cite{Van2022Entangling} &  HNA lens \cite{Rosenfeld2017Event-Ready}  \\
      \hline
      $P_{E}$   & $6 \times 10^{-3}$  & $0.1$ & $1.22 \times 10^{-6}$ &   $7 \times 10^{-7}$  \\
      rate ($s^{-1}$) & 6 & $3.2 \times 10^{3}$ & 0.005 & 0.003   \\
    \hline
    &  QA & Ions \cite{Slodiifmmode2013Atom-Atom} & Ions \cite{Saha2025High} & Ions \cite{Stephenson2020High-Rate}  \\
    \hline 
    $P_{E}$   & 0.06 & $1.1 \times 10^{-4}$ & $2.3 \times 10^{-5}$ & $2.18 \times 10^{-4}$ \\
    rate ($s^{-1}$) & $1.66 \times 10^{4}$ & 0.23 & 0.35 & 182
    \end{tabular}
    \caption{Comparison of the remote atom-atom entanglement success probability $P_E$ and generation rate for cavity, high-NA lens, trapped ions, and our quantum antenna (QA) schemes. For additional data, see Refs. \cite{Reiserer2015Cavity-based, Reiserer2022Colloquium, Wei2022Towards} and references therein. }
    \label{tab-comparison}
\end{table}

We emphasize the achievable probability and the generation rate of remote atom-atom entanglement enabled by our architecture. 
The probability of entangling the remote atom qubit is $P_\text{E} = \frac{1}{2}(\eta_\text{t} \eta_\text{d} \eta)^2$, with $\eta_\text{t}$ and $\eta_\text{d}$ being the total optical transmission and fiber coupling efficiency, and the single photon detection efficiency, respectively. The factor $\frac{1}{2}$ arises from the success probability of the Bell measurement. 
For a typical single photon detector at $780$ nm, we expect $\eta_\text{d} \simeq 0.9$ \cite{Esmaeil2021Superconducting,strohauer2025current}, and $\eta_\text{t}=0.7$, giving $P_\text{E} \simeq 6 \%$. This is comparable to cavity or high-NA-lens assisted schemes \cite{Wei2022Towards, Young2022Architecture}, as shown in Table~\ref{tab-comparison}. 

Finally, we estimate the remote entanglement generation rate, $d_r$, achieved by our scheme.
It is defined as the number of entangled communication-atom pairs generated within a duration $ \tau = 1\,\text{s} $.
In the ideal case—neglecting any laser-induced heating—the entanglement generation process consists of three main steps: (1) Atom-ensemble entanglement generation, taking approximately $ \tau_w \simeq 1\ \mu\text{s} $; (2) Retrieval of the spin-wave excitation as a single photon, with a duration of $ \tau_r \simeq 1\,\mu\text{s}$; (3) Others including electronic processing time, single atom Rydberg excitation, and the separation time between write and retrieval, estimated as $ \tau_p \simeq 1\,\mu\text{s}$.
This gives a total cycle time of $\tau_{\text{cycle}} = \tau_w + \tau_r + \tau_p \simeq 3\,\mu\text{s}$, corresponding to $N_t = \tau/\tau_{\text{cycle}} \simeq \frac{1}{3} \times 10^6$ possible cycles per second.
However, not every cycle results in successful entanglement. Given a success probability of \( P_E \simeq 6\% \), the ideal entanglement generation rate is $d_r = N_t P_E/\tau \simeq 20\,\text{kHz}$.  
In practice, continuous coupling of both the communication atom and the atomic ensemble to Rydberg states via laser fields leads to unavoidable heating.
The heating rate for the communication atom is estimated as $1/\tau_{r,\text{atom}} \sim 10\,\text{kHz}$, while the ensemble experiences much weaker heating due to far-detuned laser coupling, with a rate $\Omega^2/4\Delta^2 \tau_{r,\text{atom}} \simeq 25\,\text{Hz}$. However, the duration of single-atom excitation and retrieval is much shorter than that of the dipole–exchange interaction in the ensemble. 
To mitigate these effects, we introduce periodic state preparation every 20 cycles and cooling every 2000 cycles. Preparing the atoms back to the ground state takes about 1 $\mu \rm s$, while cooling the ensemble requires about 1 ms.
These additional cooling processes reduce the effective entanglement rate by multiplicative factors of $\frac{60}{61} $ and $\frac{6000}{7100}$, respectively.
Taking these into account, the practical entanglement generation rate becomes
\begin{equation}
 d_r \simeq 16.6 \,\text{kHz},   
\end{equation}
which significantly exceeds the rates achieved by existing schemes based on optical cavities or high-NA lenses, as shown in the Table~\ref{tab-comparison}.

\textit{Conclusion and Discussion}: 
In this work, we proposed a quantum network architecture based on a quantum antenna. By integrating neutral atoms as memory and communication qubits with an atomic ensemble serving as the antenna, our approach combines long qubit coherence times with a high entanglement generation rate of 16.6 kHz. Owing to its simplicity, tunability, and experimental accessibility, this architecture offers a promising foundation for near-term distributed quantum computing and sensing with neutral atoms.
An additional performance advantage of our architecture is the built-in quantum memory, which stores the atom–ensemble entangled state throughout photon transmission and heralding, mitigating link delays and enhancing the effective entanglement generation rate—an advantage over cavity- and high-NA-lens-based schemes that lack long-lived, trap-compatible Rydberg memories.

Several future directions arise from this work. First, the architecture can be naturally extended to include multiple communication atoms, enabling distributed, large-scale quantum computing \cite{Campbell2017Roads}. 
Second, by employing higher Rydberg states with stronger dipole interactions or using shortcuts-to-adiabaticity protocols \cite{Odelin2019Shortcuts}, the adiabatic passage time can be further reduced, thereby increasing the entanglement generation rate. Third, incorporating a bad cavity could further optimize photon generation performance with minimal added complexity.

\textit{Acknowledgment}: XL and CZ acknowledge the support of the Air Force Office of Scientific Research under Grant No. FA9550-20-1-0220 and the National Science Foundation under Grant No. PHY-2409943, OSI-2228725, ECCS-2411394.
YM acknowledges the support from the WSU New Faculty Seed Grant.

\bibliography{ref}

\end{document}


\title{Supplementary Information of “Cold atomic ensembles as quantum antennas for distributed networks of single-atom arrays”}
\author{Xiaoshui Lin}
\affiliation{Department of Physics, Washington University, St. Louis, MO, 63130, USA}

\author{Yefeng Mei}
\affiliation{Department of Physics and Astronomy, Washington State University, Pullman, Washington 99164, USA}

\author{Chuanwei Zhang}
\email{chuanwei.zhang@wustl.edu}
\affiliation{Department of Physics, Washington University, St. Louis, MO, 63130, USA}
\affiliation{Department of Physics, University of Texas at Dallas, Richardson, TX, 75080, USA}

\date{\today}

\maketitle 

\tableofcontents

\section{Entangling the communication atom with the atomic ensemble}
\label{sec-1}

Here, we present the process of entangling the communication atom with the atomic ensemble. 
The key ingredient for this process is the dipole-dipole interaction between the communication atom and the atomic ensemble. 
To better clarify the dynamics and the Hamiltonian, we start with a simple setup that entangles the communication atom with the ensemble using one set of Rydberg states. 
For this setup, the entangled state between the communication atom and the ensemble is
\begin{equation}
|\psi \rangle  = \frac{1}{\sqrt{2}}(|\downarrow \rangle |G \rangle + |\uparrow \rangle |S \rangle.
\end{equation}
where $|G \rangle = |g\rangle_1 |g \rangle_2 \dots |g \rangle_N$ and $|S\rangle = \sum_j C_j e^{i \mathbf{k}\cdot \mathbf{r}_j} |g\rangle_1 |g \rangle_2 \dots |s\rangle_j\dots |g \rangle_N$. 
Then, we clarify how we can establish entanglement between the communication atom and the atomic ensemble using two Rydberg states. 
In this case, the wave function is 
\begin{equation}
|\psi \rangle  = \frac{1}{\sqrt{2}}(|\downarrow \rangle |S_{\downarrow} \rangle + |\uparrow \rangle |S_{\uparrow} \rangle.
\end{equation}
Using this wave function and dual-rail scheme, we can further entangle the communication atom with a polarization photon qubit.

\subsection{Single Rydberg state scheme}

\begin{figure}[htbp!]
    \centering
    \includegraphics[width=0.35 \linewidth]{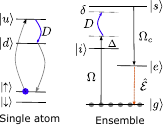}
    \caption{ Energy levels for the single atoms and the atomic ensemble. 
    We use the lower hyperfine state of the atom as the memory qubits, denoting $| \uparrow \rangle$ and $|\downarrow \rangle$. 
    The state $|\uparrow \rangle$ can be coupled to the Rydberg state $|u\rangle$. 
    The transition $|u \rangle \rightarrow | d\rangle$ is coupled to the atomic ensemble's Rydberg states $| s \rangle$ and $| i \rangle$, through dipole-dipole exchange interaction $D(\mathbf{r} - \mathbf{r}_s)$.
    A laser pulse couples the atomic ensemble's ground state $|g \rangle$ to the intermediate state $|i \rangle$ with Rabi frequency $\Omega$ and detuning $\Delta \gg \Omega$.  
    Together with the dipole-dipole exchange $|u \rangle |i \rangle$ to $|d \rangle |s \rangle$, this leads to a transition $|g \rangle \rightarrow |s \rangle$ detuned by $\delta$.
    The state $|s\rangle$ can be further resonantly coupled to state $|e\rangle$ and emits a single photon $\hat{\mathcal{E}}$ and back to the ground state $|g\rangle$. }
    \label{supfig1}
\end{figure}

For the single Rydberg state scheme, the relevant energy levels are presented in Fig. \ref{supfig1}. 
For the communication atom, the two lower states are the memory qubit.
Initially, it is prepared in a superposition state $\frac{1}{\sqrt{2}} (| \downarrow \rangle + | \uparrow \rangle)$ using a $\frac{\pi}{2}$ pulse. 
Then, we pump the state $|\uparrow \rangle$ to the Rydberg state $|u \rangle$ for interacting. 
Once the communication atom is pumped into the Rydberg state, we couple the ensemble from the ground state $|g \rangle$ to an intermediate Rydberg state $|i \rangle$ with a large detuning $\Delta$.
As Rydberg states, the paired state $|u \rangle|i \rangle$ can be flipped into another paired state $|d\rangle|s \rangle$ via a dipole-dipole exchange interaction $D_j = \frac{C_3}{2|\mathbf{r}_j - \mathbf{r}_c|^3}[1 - 3 \cos^2(\theta_j)]$.
Here $\mathbf{r}_j$ is the position of the $j$-th atom in the ensemble, $\mathbf{r}_c$ is the position of the communication atom, and $\theta_j$ is the angle between the quantized axis and $\mathbf{r}_j - \mathbf{r}_c$. 
We first consider the case where there is only one atom in the ensemble.
The following Hamiltonian \ captures their dynamics \begin{align}
    H &= \left(\omega_u | u\rangle \langle u | + \omega_d | d\rangle \langle d |\right) \otimes I_\mathrm{R} + I_\mathrm{S} \otimes \left(\omega_s | s\rangle \langle s | + \omega_i | i \rangle \langle i | + \omega_g | g \rangle \langle g | \right) \nonumber \\
    &+ \left( D | u \rangle \langle d | \otimes | i \rangle \langle s | + \Omega(t) I_\text{S} \otimes |i \rangle\langle g| +  \mathrm{H.c.}
    \right).   
\end{align} 
Using the basis $\{|u,s\rangle, |u, i\rangle, |u, g\rangle, |d, s\rangle, |d,i\rangle, |d, g\rangle\}$, the above Hamiltonian is rewritten into a matrix form as 
\begin{equation}
    H = \begin{pmatrix}
        \omega_u + \omega_s & 0 & 0 & 0 & 0 & 0 \\
        0 & \omega_u + \omega_i & \Omega(t) & D & 0 & 0  \\        
        0 & \Omega(t) & \omega_u + \omega_g & 0 & 0 & 0\\        
        0 & D & 0 & \omega_d + \omega_s & 0 & 0 \\ 
        0 & 0 & 0 & 0 & \omega_d + \omega_i  & \Omega(t) \\ 
        0 & 0 & 0 & 0 & \Omega(t) & \omega_d + \omega_g  \\
    \end{pmatrix} \begin{array}{c}
         |u,s\rangle \\
         |u, i\rangle \\
         |u, g\rangle \\
         |d, s\rangle \\ 
         |d,i\rangle \\
         |d, g\rangle 
    \end{array}.
    \label{eq-unitary-transform}
\end{equation}
Using the following rotating operator 
\begin{equation}
    U = \left( |u \rangle \langle u | \rangle + |d\rangle\langle d| \right)\otimes \left(|s \rangle \langle s | \rangle + |i\rangle \langle i| + \mathrm{e}^{-\mathrm{i} \omega t } |g \rangle \langle g|  \right),
\end{equation}
the effective Hamiltonian writes as 
\begin{align}
    H_\text{eff} &= U H U^{\dagger} - \mathrm{i} U \partial_t U^{\dagger} \nonumber \\ 
    & = \begin{pmatrix}
        \omega_s + \omega_u & 0 & 0 & 0 & 0 & 0 \\
        0 & \omega_i + \omega_u & \Omega(t)e^{i  \omega t} & D & 0 & 0  \\        
        0 & \Omega(t) e^{-i \omega t} & \omega + \omega_g + \omega_u & 0 & 0 & 0\\        
        0 & D & 0 & \omega_d + \omega_s & 0 & 0 \\ 
        0 & 0 & 0 & 0 & \omega_d + \omega_i  & \Omega(t)e^{i \omega t} \\ 
        0 & 0 & 0 & 0 & \Omega(t) e^{-i \omega t} & \omega +  \omega_d + \omega_g  \\
    \end{pmatrix} \begin{array}{c}
         |u,s\rangle \\
         |u, i\rangle \\
         |u, g\rangle \\
         |d, s\rangle \\ 
         |d,i\rangle \\
         |d, g\rangle 
    \end{array}.
\end{align}
Taking $\Omega(t) = 2\Omega \cos(\omega t)$ and neglecting anti-rotating wave terms, we find 
\begin{align}
    H_\text{eff} = \left( \begin{array}{c|ccc|cc}
        \omega_s - \omega_u & 0 & 0 & 0 & 0 & 0 \\
        \hline 
        0 & \omega_i +\omega_u & \Omega & D & 0 & 0  \\        
        0 & \Omega & \omega + \omega_g + \omega_u & 0 & 0 & 0\\        
        0 & D & 0 & \omega_d + \omega_s   & 0 & 0 \\ 
        \hline
        0 & 0 & 0 & 0 & \omega_d + \omega_i  & \Omega \\ 
        0 & 0 & 0 & 0 & \Omega & \omega + \omega_d + \omega_g 
    \end{array} \right) \begin{array}{c}
         |u,s\rangle \\
         |u, i\rangle \\
         |u, g\rangle \\
         |d, s\rangle \\ 
         |d,i\rangle \\
         |d, g\rangle 
    \end{array}.
\end{align}
Based on this Hamiltonian, the Hilbert space can be decoupled into three subspaces.
If the atom is in $|u\rangle$ and the ensemble is in ground state $|g\rangle$, the relevant subspace is expanded by the basis $|u, i\rangle$, $|u, g\rangle$, and $|d, s\rangle$. 
In this subspace, the effective Hamiltonian writes as
\begin{align}
    H_\text{eff} &=  [(\omega_i  - \omega_g) - \omega ]  |u \rangle\langle u| \otimes |i \rangle\langle i | + 0 \times |u \rangle\langle u| \otimes |g \rangle\langle g | + [(\omega_d - \omega_u) - \omega + (\omega_s - \omega_g )] |d \rangle\langle d| \otimes |s \rangle\langle s | \nonumber \\
    &+ (D |u\rangle \langle d| \otimes |i \rangle \langle s|  + \Omega |u \rangle \langle u |\otimes |i \rangle \langle g|+ \mathrm{H.c.}).
\end{align}
Defining $\Delta = \omega - (\omega_i - \omega_g)$ and  $\delta = \omega + (\omega_u - \omega_d) - (\omega_s - \omega_g)$, it is reduced to
\begin{align}
    H_\text{eff} &= -\Delta |u \rangle\langle u| \otimes |i \rangle\langle i | -\delta  |d \rangle\langle d| \otimes |s \rangle\langle s | + (D |u\rangle \langle d| \otimes |i \rangle \langle s|  + \Omega  |u \rangle \langle u |\otimes |i \rangle \langle g|+ \mathrm{H.c.}). 
    \label{eq-effective-three-level}
\end{align}
Here, we consider that there are $N \gg 1$ atoms in the ensemble and the coupling has spatial dependence as $\Omega e^{i \mathbf{k}_0 \cdot \mathbf{r}_j}$. 
It leads to the effective Hamiltonian $\Tilde{H}_1$ as 
\begin{align}
    \tilde{H} =& - \sum_j \left\{ \Delta |u \rangle_s \langle u| \otimes |i \rangle_j \langle i | + \delta  |d \rangle_s \langle d| \otimes |s \rangle_j \langle s |  - [ D_j |u\rangle_s \langle d| \otimes |i \rangle_j \langle s| + \Omega e^{i \mathbf{k}_0 \cdot \mathbf{r}_j}|u \rangle_s \langle u | \otimes |i \rangle_j \langle g|+ \mathrm{H.c.}]  \right\}, 
    \label{eq-three-level-simulation}
\end{align}
where $D_j = D(\mathbf{r}_j - \mathbf{r}_c) = \frac{C_3}{2|\mathbf{r}_j - \mathbf{r}_c|^3} [1 - 3\cos^2(\theta_j)]$. 
$\theta$ is the angle between $\mathbf{r}_j - \mathbf{r}_c$ and the $z$ axis. 
Assuming a large detuning is set as  $\Delta \gg |\Omega|$, $\delta$, and $|D(\mathbf{R}_j)|$, we can adiabatically eliminate the state $|u \rangle\langle u|\otimes |i \rangle_j \langle i|$, yielding the Hamiltonian in the main text
\begin{align}
    \tilde{H}_{1,\text{eff}} &= - \sum_j \tilde{\delta}_j  |d \rangle_s \langle d| \otimes |s \rangle_j \langle s | -( \tilde{D}_j e^{i \mathbf{k}_0 \cdot \mathbf{r}_j} |u\rangle_s \langle d| \otimes |g \rangle_j \langle s| + \mathrm{H.c.}) \nonumber \\
    &= -\sum_j \begin{pmatrix}
        0 & -\tilde{D}_j e^{i \mathbf{k}\cdot \mathbf{r}_j} \\
        -\tilde{D}_j^{*} e^{-i \mathbf{k}\cdot \mathbf{r}_j} & \tilde{\delta}_j
    \end{pmatrix},
    \label{eq-two-level-simulation}
\end{align}
where $\tilde{\delta}_j = \delta + \frac{|\Omega|^2 - |D(\mathbf{R}_j)|^2}{\Delta} $ and $\tilde{D}_j = \frac{\Omega D(\mathbf{R}_j)}{\Delta}$. There is a set of zero energy states as $|u \rangle_s \langle u| \otimes |g \rangle_j \langle g | $.
We further define a collective spin-wave state as 
\begin{equation}
    |S \rangle = \frac{1}{\bar{D}} \sum_j\tilde{D}_j e^{i \mathbf{k}\cdot \mathbf{r}_j} |s\rangle_j,
\end{equation}
where $\bar{D} = \sqrt{\sum_j |\tilde{D}_j|^2}$ and neglect the difference of the detuning $\delta_j$ as $\tilde{\delta}$.
Defining $|G \rangle = |g\rangle_1 \otimes |g\rangle_2 \otimes \dots \otimes |g\rangle_N $, the Hamiltonian can be rewritten on this basis as 
\begin{equation}
    \tilde{H}_1 = -\tilde{\delta} | u \rangle \langle u| \otimes |S \rangle \langle S| + \overline{D} (|u\rangle \langle d| \otimes |G \rangle \langle S | + \mathrm{H.c.}),
\end{equation}
where only two states are relevant. The other energy levels can be viewed as dark states. 
The adiabatic dynamics of this two-level model have been intensively studied.
The eigenstates and eigenvalues of this model are
\begin{align}
    &E_{\pm} = -\frac{\tilde{\delta}}{2} \pm \sqrt{\bar{D}^2 + \frac{\tilde{\delta}^2}{4}}, \nonumber \\
    &|\psi_{\pm} \rangle =  \sqrt{\frac{1}{2}(1 \pm \frac{\delta/2}{\sqrt{\bar{D}^2 + \delta^2/4}})}| u \rangle\otimes |G\rangle \pm \sqrt{\frac{1}{2}(1 \mp \frac{\delta/2}{\sqrt{\bar{D}^2 + \delta^2/4}})} | d \rangle\otimes |S\rangle.
\end{align}
When $\delta \rightarrow -\infty$ ($\delta \rightarrow \infty$), the eigenstates are $|\psi_{+} \rangle \sim | d \rangle\otimes |S\rangle$ and $|\psi_{-} \rangle \sim | u \rangle\otimes |G\rangle$ ($|\psi_{-} \rangle \sim | d \rangle\otimes |S\rangle$ and $|\psi_{+} \rangle \sim | u \rangle\otimes |G\rangle$).
Therefore, the Rydberg atom spin-wave excitation state $|S\rangle$ can be adiabatically prepared by letting $\tilde{\delta} = \alpha t $ and the initial state as $|u\rangle \otimes |G\rangle$. As long as the speed $\alpha$ is small, the non-adiabatic Landua-Zeno transition is $P_{|\psi_{-} \rangle \rightarrow | \psi_{+} \rangle} = e^{- 2 \pi \overline{D}^2/ \alpha }$, which indicate a state transfer from $|u \rangle \otimes |G \rangle$ to $|d \rangle \otimes |S \rangle$. 
Then, we use a short pulse to pump the communication atom back to state $|\uparrow \rangle$ from state $|d\rangle$.
Thus, based on the above analysis, if the atom is in state $|\downarrow \rangle$, it will not be coupled to the ensemble, and the resulting final state is $| \downarrow \rangle |G \rangle$.
In contrast, if the atom is initially in state $|\uparrow \rangle$, it will be pumped into the Rydberg state, causing the state of the ensemble to become $|S \rangle$.
Thus, if the initial state of the communication atom is a superposition state as $(|\downarrow \rangle + |\uparrow \rangle )/\sqrt{2}$, the resulting wave function is as 
\begin{equation}
    |\psi\rangle = \frac{1}{\sqrt{2}} (|\downarrow \rangle |G \rangle +  e^{i \phi_d }|\uparrow \rangle |S \rangle),
    \label{eq-atom-ensemble-entangle-si}
\end{equation}
where $\phi_d$ is the dynamical phase. 
The above analysis demonstrates that our proposed scheme can generate entanglement between a single atom and the atomic ensemble.

\subsection{Two Rydberg states scheme}

\begin{figure}[!htbp]
    \centering
    \includegraphics[width=0.58\linewidth]{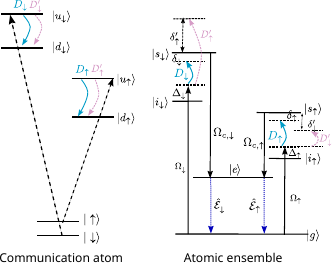}
    \caption{
    Energy levels for the communication atom and atomic ensemble.
    Two sets of individual energy levels are used to convert the communication atom state into the spin-wave state $|s_{\uparrow} \rangle$ or $|s_{\downarrow} \rangle$. 
    The dynamical process for preparing each spin-wave state is designed to be independent.
    Once all the spin-wave states are prepared, we use a short pulse to flip the communication atom back into its initial state. 
    Then, we couple the spin-wave states  $|s_{\downarrow}\rangle$ and $|s_{\uparrow}\rangle$ to the intermediate state $|e\rangle$.
    It will decay to the ground state and emit a photon.
    }
    \label{supfig2}
\end{figure}

With the above analysis, it is straightforward to construct a scheme using two Rydberg states. 
The energy levels are presented in Fig. \ref{supfig2}.
Here, the two lower states $| \downarrow\rangle$ and $| \uparrow\rangle$ are sperately pumped into the Rydberg state $|u_{\downarrow} \rangle$ and $|u_{\uparrow} \rangle$. 
Then, we couple the ground state $|g\rangle$ of the ensemble to intermediate Rydberg state $|i_{\downarrow} \rangle$ with detuning $\Delta_{\downarrow}$ and Rydberg state $|i_{\uparrow} \rangle$ with detuning $\Delta_{\uparrow}$. 
Then, the paired state $|u_{\downarrow} \rangle |i_{\downarrow} \rangle$ will interact with paired state $|d_{\downarrow} \rangle |s_{\downarrow} \rangle$.
Different from what we have discussed in Sec. \ref{sec-1}, the communication atom will also interact with the ensemble via interaction between paired states $|u_{\downarrow} \rangle |i_{\uparrow} \rangle$ and  $|d_{\downarrow} \rangle |s_{\uparrow} \rangle$.
We denote the strength of the first kind interaction as $D_{\downarrow} $ and the second kind as $D_{\downarrow}'$. 
The same issue will also appear for the paired state $|u_{\uparrow} \rangle |i_{\uparrow} \rangle$ ,and we denote the strength here as $D_{\uparrow} $ and $D_{\uparrow}'$.
As a starting point, we consider that there is only one atom in the ensembles.
The total Hilbert space dimension is $d = 4 \times 5 = 20$, which is too complicated to solve in general conditions. 
However, since the dipole-dipole exchange interaction conserves the total excitation, the Hilbert space will be decoupled into several subspaces. 
For initial states $|u_{\downarrow} \rangle|G \rangle$ and $|u_{\uparrow} \rangle|G \rangle$, the effective Hamiltonian in the rotating frame for these two subspaces can be written as 
\begin{equation}
    H_{\text{eff},\downarrow} = \begin{pmatrix}
        -\Delta_{\downarrow} & 0 & \Omega_{\downarrow} & D_{\downarrow} & 0  \\
        0 & -\Delta_{\uparrow} & \Omega_{\uparrow} & 0 & D_{\downarrow}'  \\        
        \Omega_{\downarrow} & \Omega_{\uparrow} & 0 & 0 & 0\\        
        D_{\downarrow} & 0 & 0 & -\delta_{\downarrow} & 0  \\ 
        0   &D_{\downarrow}' & 0 & 0& -\delta_{\downarrow}' 
    \end{pmatrix} \begin{array}{c}
         |u_{\downarrow}, i_{\downarrow}\rangle \\
         |u_{\downarrow}, i_{\uparrow}\rangle \\
         |u_{\downarrow}, g\rangle \\
         |d_{\downarrow}, s_{\downarrow}\rangle \\ 
         |d_{\downarrow}, s_{\uparrow}\rangle 
    \end{array},
    \label{eq-effect-1}
\end{equation}
and 
\begin{equation}
    H_{\text{eff},\uparrow} = \begin{pmatrix}
        -\Delta_{\downarrow} & 0 & \Omega_{\downarrow} & D_{\uparrow}' & 0  \\
        0 & -\Delta_{\uparrow} & \Omega_{\uparrow} & 0 & D_{\uparrow}  \\        
        \Omega_{\downarrow} & \Omega_{\uparrow} & 0 & 0 & 0\\        
        D_{\uparrow}' & 0 & 0 & -\delta_{\uparrow}' & 0  \\ 
        0   &D_{\uparrow} & 0 & 0& -\delta_{\uparrow} 
    \end{pmatrix} \begin{array}{c}
         |u_{\uparrow}, i_{\downarrow}\rangle \\
         |u_{\uparrow}, i_{\uparrow}\rangle \\
         |u_{\uparrow}, g\rangle \\
         |d_{\uparrow}, s_{\downarrow}\rangle \\ 
         |d_{\uparrow}, s_{\uparrow}\rangle 
    \end{array}, 
\end{equation}
with $\Delta_{\downarrow} = \omega_{\downarrow} - \omega_{ig, \downarrow}$, $\Delta_{\uparrow} = \omega_{\uparrow} - \omega_{ig, \uparrow} $, $\delta_{\downarrow} = \omega_{\downarrow} + \omega_{ud, \downarrow} - \omega_{sg, \downarrow}$, $\delta_{\downarrow}' = \omega_{\uparrow} + \omega_{ud, \downarrow} - \omega_{sg, \uparrow} = \Delta_{\uparrow} + \omega_{ud,\downarrow} - \omega_{si,\uparrow}$,
$\delta_{\uparrow} = \omega_{\uparrow} + \omega_{ud, \uparrow} - \omega_{sg, \uparrow}$, $\delta_{\uparrow}' = \Delta_{\downarrow} + \omega_{ud,\uparrow} - \omega_{si,\downarrow}$. 
Apparently, if $D_{\downarrow}'$ and $D_{\uparrow}'$ are zero, the above effective Hamiltonian reduces to the Hamiltonian we studied in the first section. However, they are not zero in general conditions. 
To get more insight into the Hamiltonian and associated dynamics, we focus on the dynamics in the first subspace and rearrange the basis to make the matrix in Eq. \eqref{eq-effect-1} as 
\begin{align}
    H_{\text{eff},\downarrow} = \left( \begin{array}{ccc|cc}
        -\Delta_{\downarrow} & 0 & 0 & \Omega_{\downarrow} & D_{\downarrow}   \\        
        0 & -\Delta_{\uparrow} & D_{\downarrow}' & \Omega_{\uparrow} & 0 \\        
        0 & D_{\downarrow}' & -\delta_{\downarrow}' & 0 & 0 \\ 
        \hline
        \Omega_{\downarrow} & \Omega_{\uparrow} & 0 & 0 & 0\\ 
        D_{\downarrow} & 0 & 0 & 0 & -\delta_{\downarrow}
    \end{array} \right) \begin{array}{c}
         |u_{\downarrow}, i_{\downarrow}\rangle \\
         |u_{\downarrow}, i_{\uparrow}\rangle \\
         |d_{\downarrow}, s_{\uparrow}\rangle \\
         |u_{\downarrow}, g\rangle \\
         |d_{\downarrow}, s_{\downarrow}\rangle
    \end{array}.
\end{align}
We employ a local transformation on subspace expanded by $|u_{\downarrow}, i_{\uparrow} \rangle$ and $|d_{\downarrow}, s_{\uparrow} \rangle$. We define two new basis $|\psi_{+} \rangle = [(\lambda -a ) |u_{\downarrow}, i_{\uparrow} \rangle + b |d_{\downarrow}, s_{\uparrow} \rangle]/\sqrt{2\lambda(\lambda - a)}$ and $|\psi_{-} \rangle = [-b |u_{\downarrow}, i_{\uparrow} \rangle + (\lambda - a ) |d_{\downarrow}, s_{\uparrow} \rangle]/\sqrt{2\lambda(\lambda - a)}$, where  $a = (\Delta_{\uparrow} - \delta_{\downarrow}')/2 = -(\omega_{ud,\downarrow} - \omega_{si,\uparrow})/2 $, $b = D_{\downarrow}'$, and $\lambda = \sqrt{a^2 + b^2}$.  
The Hamiltonian in this new basis becomes
\begin{align}
    H_{\text{eff},\downarrow} = \left( \begin{array}{ccc|cc}
        -\Delta_{\downarrow} & 0 & 0 & \Omega_{\downarrow} & D_{\downarrow}   \\        
        0 & -\frac{\Delta_{\uparrow} + \delta_{\downarrow}'}{2} - \lambda & 0 & -\Omega_{\uparrow} \frac{b}{\sqrt{2\lambda(\lambda - a)}} & 0 \\        
        0 & 0 & -\frac{\Delta_{\uparrow} + \delta_{\downarrow}'}{2} +  \lambda & \Omega_{\uparrow} \sqrt{\frac{1}{2}(1 - \frac{a}{\lambda})} & 0 \\ 
        \hline
        \Omega_{\downarrow} & -\Omega_{\uparrow} \frac{b}{\sqrt{2\lambda(\lambda - a)}}  & \Omega_{\uparrow} \sqrt{\frac{1}{2}(1 - \frac{a}{\lambda})} & 0 & 0\\ 
        D_{\downarrow} & 0 & 0 & 0 & -\delta_{\downarrow}
    \end{array} \right) \begin{array}{c}
         |u_{\downarrow}, i_{\downarrow}\rangle \\
         |\psi_{-} \rangle \\
         |\psi_{+} \rangle \\
         |u_{\downarrow}, g\rangle \\
         |d_{\downarrow}, s_{\downarrow}\rangle
    \end{array}.
    \label{eq-rotating}
\end{align}
Here, we have not made any approximation, and the results are exact.
In the experiments, we always set $\Delta_{\downarrow}$ and $-\frac{\Delta_{\uparrow} + \delta_{\downarrow}'}{2} \pm \lambda$ are much larger than the other paramters.
Thus, we can eliminate the states $|u_{\downarrow}, i_{\downarrow}\rangle$, $|\psi_{-} \rangle$, and $|\psi_{+} \rangle$. 
The resulting effective Hamiltonian for states $|u_{\downarrow}, g\rangle$ and $|d_{\downarrow}, s_{\downarrow}\rangle$ is
\begin{equation}
    \tilde{H} = \begin{pmatrix}
    \frac{\Omega_{\downarrow}^2}{\Delta_{\downarrow}} + \Omega_{\uparrow}^2 \frac{ b^2 (\lambda - \lambda_0) + (\lambda - a)^2(\lambda + \lambda_0)}{2 (\lambda^2 - \lambda_0^2) \lambda (\lambda - a)} & \frac{\Omega_{\downarrow} D_{\downarrow}}{\Delta_{\downarrow}}  \\
   \frac{\Omega_{\downarrow} D_{\downarrow}}{\Delta_{\downarrow}}    & - \delta_{\downarrow} + \frac{D_{\downarrow}^2}{\Delta_{\downarrow}} 
      
    \end{pmatrix},
\end{equation}
with $\lambda_0 = \frac{\Delta_{\uparrow} + \delta_{\uparrow}'}{2}$. 
In our design, the value of dipole-dipole interaction is  $b = D_{\uparrow}' \sim 2\pi \times 2 \ \text{MHz}$, the value of detuning $a = (\Delta_{\uparrow} - \delta_{\uparrow}')/2 \sim  2\pi \times 100 $ MHz, and the value of $\lambda_0 \sim 2\pi \times 100$ MHz. 
This leads to 
\begin{equation}
    \lambda - a \sim \frac{b^2}{2 a}, \quad \lambda + \lambda_0 \sim \Delta_{\uparrow}  + \frac{b^2}{2 a}, \quad \lambda - \lambda_0 \sim \delta_{\uparrow}'  + \frac{b^2}{2 a} 
\end{equation} 
Thus, we may approximate the effective Hamiltonian as
\begin{equation}
       \tilde{H} \sim \begin{pmatrix}
    \frac{\Omega_{\downarrow}^2}{\Delta_{\downarrow}}  & \frac{\Omega_{\downarrow} D_{\downarrow}}{\Delta_{\downarrow}}  \\
   \frac{\Omega_{\downarrow} D_{\downarrow}}{\Delta_{\downarrow}}   & -\delta_{\downarrow} + \frac{D_{\downarrow}^2}{\Delta_{\downarrow}}, 
    \end{pmatrix} 
\end{equation}
or 
\begin{align}
    \tilde{H}_{\downarrow} &= - \sum_j \tilde{\delta}_{\downarrow,j}  |d_{\downarrow} \rangle \langle d_{\downarrow}| \otimes |s_{\downarrow} \rangle_j \langle s_{\downarrow} | -( \tilde{D}_{\downarrow,j} e^{i \mathbf{k}_{\downarrow,0} \cdot \mathbf{r}_j} |d_{\downarrow} \rangle \langle u_{\downarrow}| \otimes |g \rangle_j \langle s_{\downarrow}| + \mathrm{H.c.}),
\end{align}
with $\delta_{\downarrow,j} = \delta_{\downarrow} + \frac{\Omega_{\downarrow}^2 - D_{\downarrow}^2}{\Delta_{\downarrow}} + \frac{\Omega_{\uparrow}^2}{\Delta_{\uparrow}} $ and $\tilde{D}_j = {\Omega_{\downarrow} D_{\downarrow,j} \over \Delta_{\downarrow}}$.
This two-level effective Hamiltonian is the same as we obtained in Eq. \eqref{eq-two-level-simulation}, except the effective detuning is different.
The same result can be obtained for the subspace of $| u_{\uparrow}\rangle |g \rangle$, from which we have 
\begin{align}
    \tilde{H}_{\uparrow} &= - \sum_j \tilde{\delta}_{\uparrow,j}  | d_{\uparrow} \rangle \langle d_{\uparrow}| \otimes |s_{\uparrow} \rangle_j \langle s_{\uparrow} | -( \tilde{D}_{\uparrow,j} e^{i \mathbf{k}_{\uparrow,0} \cdot \mathbf{r}_j} |d_{\uparrow} \rangle \langle u_{\uparrow}| \otimes |g \rangle_j \langle s_{\uparrow}| + \mathrm{H.c.}), 
\end{align}
Adiabatically change the two-photon detuning $\delta_{\downarrow}$ and $\delta_{\uparrow}$, we can convert state $|\downarrow \rangle$ into $|u_{\downarrow} \rangle |S_{\downarrow} \rangle$ and $|\uparrow \rangle$ into $|u_{\uparrow} \rangle |S_{\downarrow} \rangle$.
If the initial state of the communication atom is in state $(|\uparrow \rangle + |\downarrow \rangle)/\sqrt{2}$, the wave function of the composite system is
\begin{equation}
|\Psi \rangle = \frac{1}{\sqrt{2}} (|\uparrow\rangle | S_{\uparrow}\rangle + e^{i \phi}|\downarrow\rangle |S_{\downarrow} \rangle).
\end{equation}
Here $\phi$ is the relative phase. 
We emphasize here that the entangling process of the communication atom and the ensemble is adiabatic, whose fidelity can be nearly $100 \%$.

\section{Converting the spin-wave excitation to a propagating photon}

Once the entanglement between the communication atom and the ensemble is established, we can further convert the spin-wave state into a single photon and generate entanglement between the communication atom and the photon. 
In this process, we use a dual-rail scheme for implementing polarization encoding of photons \cite{Wang2019Efficient}.  
First, the spin-wave state $|s_{\downarrow} \rangle$ is resonantly coupled to a lower intermediate state $|e \rangle$ with  with Rabi frequency $\Omega_{\downarrow,c}$ and wave vector $\mathbf{k}_{\downarrow,c}$.
Then the intermediate state $|e \rangle$ will decay to the ground state and emit a single photon along the direction $\mathbf{k}_{\downarrow} =\mathbf{k}_{\downarrow, c} - \mathbf{k}_{\downarrow, 0}$.
After a specifically designed period, we switch on the coupling $\Omega_{\uparrow,c}$ to resonantly couple the spin-wave state $|s_{\uparrow}\rangle$ to the intermediate state $|e\rangle$.
Similarly, it decays to the ground state and emits a photon propagating along direction $\mathbf{k}_{\uparrow} =\mathbf{k}_{\uparrow, c} - \mathbf{k}_{\uparrow, 0}$.
These photons are designed to be $\sigma^{+}$ polarized.
These photons are collected via a lens and further passed through a quarter-wave plate to be converted into horizontally polarized.
For the photon with spatial mode CH$_{\downarrow}$, it will pass through a half-wave plate to be converted into a vertically polarized state.

Since we set the condition $\Omega_{\downarrow,c}(t) \Omega_{\uparrow,c}(t) = 0$ all the time, the retrival process for spin-wave state $|S_{\downarrow} \rangle$ and $|S_{\uparrow} \rangle$ is independent. 
Thus, we can only present one process, and the other one is the same.
We set the quantized field, which couples the states $|e\rangle$ and $|g\rangle$, as 
\begin{equation}
    \hat{\mathcal{E}}(z, t) = \sum_{k} \hat{a}_{k}(t) \mathrm{e}^{\mathrm{i}(k - \omega_{eg}/c) \cdot z} e^{i \nu t},
\end{equation}
and assume the spin operators that are averaged over a small interval $[z - \Delta z/2, z+\Delta z/2]$ as 
\begin{equation}
    \sigma_{\mu \nu}(z) = \frac{1}{N_z} \sum_k^{N_z} | \mu \rangle_k \langle \nu|,
\end{equation}
with $[\hat{\sigma}_{\mu \nu}(z), \hat{\sigma}_{\nu'\mu'}(z')] = \frac{\hat{\sigma}_{\mu \mu'}(z)}{n(z)} \delta_{\nu  \nu' }\delta(z - z')$, $n(z)$ the density of the ensemble.  
We may obtain the equation of motion for this process as
\begin{align}
    (\partial_t + c\partial_z) \hat{\mathcal{E}}(z, t) &= i g \sqrt{N} \hat{P}(z, t), \nonumber \\
    \partial_t \hat{P}(z, t) &= - \gamma \hat{P}(z, t) + i g \sqrt{N} \hat{\mathcal{E}}(z, t) (\hat{\sigma}_{gg} - \hat{\sigma}_{ee}) +i \Omega_c \hat{S}(z, t) + \sqrt{2\gamma} F_P(t),   \nonumber \\
    \partial_t \hat{S}(z, t) &= - \gamma_s \hat{S}(z, t) - i g \hat{\mathcal{E}}(z, t) \sqrt{N} \hat{\sigma}_{es} +i  \Omega_c \hat{P}(z, t) + + \sqrt{2\gamma_s} F_S(t), 
\end{align}
with polarization $\hat{P}(z) = \sqrt{N} \hat{\sigma_{ge}} $, spin-wave operator $\hat{S} = \sqrt{N} \hat{\sigma_{gs}}  $, $F_P(t)$ and $F_{S}(t)$ is the uncorrelated Langevin noise.  
For a single photon excitation, most of the atoms are in the ground state. Thus, we can assume $ \langle \hat{\sigma}_{gg} \rangle \sim 1 $ and $\langle \hat{\sigma}_{ee} \rangle, \  \langle \hat{\sigma}_{es} \rangle \sim 0$ and the above equations becomes solvable as
\begin{align}
    (\partial_t + c\partial_z) \hat{\mathcal{E}}(z, t) &= i g \sqrt{N} \hat{P}(z, t), \nonumber \\
    \partial_t \hat{P}(z, t) &= - \gamma \hat{P}(z, t) + i g \sqrt{N} \hat{\mathcal{E}}(z, t) +i \Omega_c \hat{S}(z, t) + \sqrt{2\gamma} F_P(t),   \nonumber \\
    \partial_t \hat{S}(z, t) &= - \gamma_s \hat{S}(z, t)  + i  \Omega_c \hat{P}(z, t) + + \sqrt{2\gamma_s} F_S(t), 
\end{align}
If we neglect the Langevin noise and decay, the above equations can be further simplified.
These simplified equations may lead to the following equations for $\hat{\mathcal{E}}(z, t)$ and $\hat{S}(z, t)$
\begin{equation}
   (\partial_t + c \partial_z) \hat{\mathcal{E}}(z, t) = \frac{g \sqrt{N} }{\Omega_c(t)} \partial_t \hat{S}(z, t),
\end{equation}
which can be solved using a dark-state polarization operator 
\begin{equation}
    \hat{\Psi}(z, t) = \cos(\theta(t)) \hat{\mathcal{E}}(z, t) - \sin(\theta(t)) \hat{S}(z, t),   
\end{equation}
where 
\begin{equation}
    \cos(\theta) = \frac{\Omega_c(t)}{\sqrt{[\Omega_c(t)]^2 + g^2 N}}, \quad \sin(\theta) = \frac{g \sqrt{N}}{\sqrt{[\Omega_c(t)]^2 + g^2 N }} .
\end{equation}
The equation for this dark-stat polarization operator is
\begin{equation}
    [\partial_t + c \cos^2(\theta(t)) \partial_z] \hat{\Psi}(z, t) = 0. 
\end{equation}
The solution for this polarization operator is generally 
\begin{equation}
    \hat{\Psi}(z, t) = \hat{\Psi}\left(z - \int_{t_0}^{t} dt' c \cos^{2}[\theta(t')] , t = t_0\right).
\end{equation}
All the photon information can be encoded into this polarization operator using an appropriate control laser $\Omega(t)$.
Initially, the polarization operator $\hat{\Psi}(z, t)$ is the spin-wave state $S$ that is prepared in step 4.
Then we adiabatically switch on the control light $\Omega_c$ and let $\Omega_c \gg g \sqrt{N}$, generating a single photon and causing the atomic ensemble to return to its ground state again. 
Thus, we can reuse this atomic antenna multiple times using the above procedures.

With the above analysis, we find that if the ensemble is in state $|S_{\downarrow}$, it will create a single photon with state $| V \rangle$.
In contrast, if the ensemble is in state $|S_{\uparrow} \rangle$, it will create a single photon with state $| H \rangle$.
If the wave function for the composite system of the communication atom and the atomic ensemble is $(|\downarrow \rangle |S_{\downarrow} \rangle + e^{i \phi }|\uparrow \rangle |S_{\uparrow} \rangle  )/\sqrt{2}$, the final state will be
\begin{equation}
        |\psi\rangle = \frac{1}{\sqrt{2}} (|\downarrow \rangle |V \rangle +  e^{i \phi_d' }|\uparrow \rangle |H \rangle).
\end{equation}
This state is a maximally entangled state between the atom and the photon, which can be used to generate remote entanglement.

\section{Experiment parameters}
\begin{figure}[htbp!]
    \centering
    \includegraphics[width=0.98\linewidth]{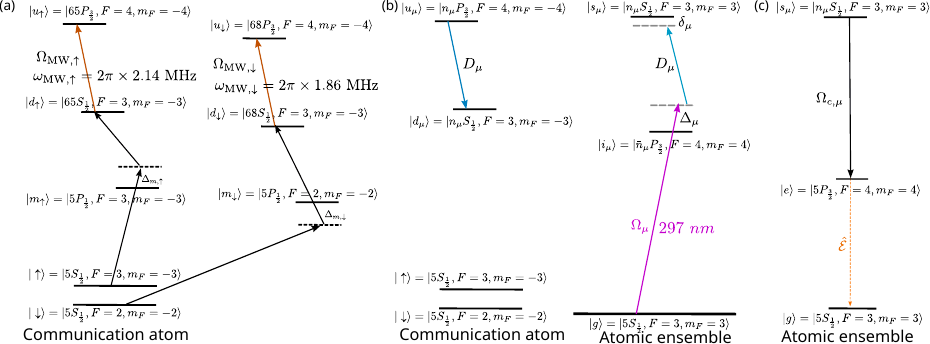}
    \caption{
    Experimental relevant level structure for $^{85}$Rb communication atom and atomic ensemble.
    For the communication atoms, the two low-lying hyperfine states are chosen as $|\uparrow \rangle = |5S_{1\over 2}, F = 3, m_F = -3\rangle$ and $|\uparrow \rangle = |5S_{1\over 2}, F = 2, m_F = -2\rangle$. 
    The single-qubit logical gate is implemented using a microwave.
    (a) Relevant energy levels for the processes $| \uparrow \rangle \rightarrow 
    |u_{\uparrow} \rangle$ and $|\downarrow \rangle \rightarrow |u_{\downarrow} \rangle$ for the communication atom. 
    We couple the state $|\uparrow \rangle$ to the Rydberg state $|d_{\uparrow} \rangle$ using a two-photon process. The single photon detuning is $\Delta_{m,\uparrow} \sim 2\pi \times 100$ MHz, and the two-photon detuning is set to be zero.
    Using a $\pi$-pulse, we can pump the communication atom from state $|\uparrow\rangle$ to $|d_{\uparrow}\rangle$ with a probability around unity.
    Then, we utilize a microwave with frequency $\omega_{\text{MW},\uparrow} = 2\pi \times 2.14$ GHz to flip state $|d_{\uparrow}\rangle$ to state $|u_{\uparrow}\rangle$ as the initial state.
    After generating the entanglement between the communication atom and the atomic ensemble, the communication atom is prepared in state $|d_{\uparrow} \rangle$. 
    We can use the same two lasers to flip the communication atom back to $|\uparrow \rangle$.
    We can utilize the same procedure for pumping state $|\downarrow \rangle$ to $|u_{\downarrow} \rangle$ with a slightly different intermediate state $|m_{\downarrow} \rangle$ and microwave frequency. 
    (b) The relevant energy levels for entangling the single atom and the atomic ensemble. 
    Here, $\mu \in\{ \uparrow, \downarrow  \}$, $n_{\uparrow} = 65$, $\bar{n}_{\uparrow} = 64$, $n_{\downarrow} = 68$, and $\bar{n}_{\downarrow} = 67$.
    We set $\Delta_{\uparrow} = 2\pi \times 147$ MHz and $\Delta_{\downarrow} = 2\pi \times 125$ MHz.
    The hybridization process between $| \uparrow \rangle$ and $|\downarrow \rangle$ is suppressed by the large detuning between each set of energy levels.
    The choice of a large magnetic quantum number can suppress the transitions outside this manifold of states. 
    (c) Relevant energy levels for retrieving the spin-wave state $|s\rangle$ to a single photon $\hat{\mathcal{E}}$.
    The intermediate state is chosen as $|e\rangle = |5P_{3\over 2}, F = 4, m_F = 4 \rangle$. Thus, it can only decay to the ground state and emit a circularly polarized photon. 
    }
    \label{supfig3}
\end{figure}

Although we have demonstrated the above theoretical proposal, many gaps still exist between the theory and the experiments.
The first one is the choice of the energy levels which satisfy $\omega_{si,\mu} - \omega_{ud,\mu} \simeq \Delta_{\mu} \simeq 100$ MHz.
The second one is how to suppress the transition outside our target state manifold.

To address the above two questions, as shown in Fig. \ref{supfig3}, we may utilize the following relevant states $|\uparrow \rangle=|5 S_{1\over2}, F = 3, m_F = -3 \rangle$, $|u_{\uparrow} \rangle=|65 P_{3\over2}, F = 4, m_F = -4 \rangle$, $|d_{\uparrow} \rangle = |65 S_{1\over2}, F = 3, m_F = -3 \rangle$,  $|s_{\uparrow} \rangle=|65 P_{1\over2}, F = 3, m_F = 3 \rangle$, $|i_{\uparrow} \rangle = |64 P_{3\over2}, F = 4, m_F = 4 \rangle$ with $C_{3,\uparrow} = 17.44$ GHz $\mu m^{3}$, and $|\downarrow \rangle=|5 S_{1\over2}, F = 2, m_F = -2 \rangle$, $|u_{\downarrow}\rangle=|68 P_{3\over2}, F = 4, m_F = - 4 \rangle$, $|d_{\downarrow} \rangle = |68 S_{1\over2}, F = 3, m_F = -3 \rangle$,  $|s_{\downarrow}\rangle=|68 P_{1\over2}, F = 3,  m_F = 3 \rangle$, $|i_{\downarrow} \rangle = |67 P_{3\over2}, F = 4,  m_F = 4 \rangle$ with $C_{3,\downarrow} = 21.08$ GHz $\mu m^{3}$, the parameters for the Hamiltonian in Eq. \eqref{eq-effect-1} are
\begin{align}
    \Delta_{\uparrow} \simeq 2\pi \times 147\ \text{MHz}, \ \Delta_{\downarrow} \simeq 2\pi \times 125\ \text{MHz}, \ &\max(\delta_{\uparrow}) \sim 2\pi \times 10\ \text{MHz}, \ 
    \max(\delta_{\downarrow}) \sim 2\pi \times 10\ \text{MHz}, \nonumber \\
    \min(\delta_{\uparrow}') \sim 2\pi \times -282\ \text{MHz}, \
    &\min(\delta_{\downarrow}') \sim 2\pi \times 282\ \text{MHz}.
\end{align}
Since we use the state with a large magnetic quantum number $m_F$, the transition to states with the same energy but different magnetic quantum numbers $m_F$ is forbidden.
Meanwhile, the transition to the other manifold is suppressed by the energy mismatch.
For the atomic ensemble, we suppose the density distribution of the atoms to be a Gaussian function along the z, y, and x direction
\begin{equation}
    \rho(x,y,z) = \frac{8 N }{(2\pi)^{3 \over 2}\sigma_{\bot}^2 \sigma_z} \exp(-\frac{2 z^2}{ \sigma_z^2} - \frac{2 (x^2 + y^2)}{ \sigma_{\bot}^2}),
\end{equation}
with $\sigma_x = 4\ \mu m$, $\sigma_z = 4 \ \mu m$, and $\sigma_z = 30\ \mu m$.  If we chose the particle number $N = 500$, we get the peak density as $\rho_\text{peak} = 0.53\ \mu m^{-3}$. 
We place the communication atom at the position $(x, y, z) \simeq (7.1, 0, 0)$ $\mu m$.
The interaction between the communication atom and the atom in the ensemble center is given as
\begin{equation}
     \text{center}(D_{\uparrow, j}) = 2\pi \times 3.88 \ \text{MHz}, \quad \text{center}(D_{\downarrow, j}) =  2\pi \times 4.69\ \text{MHz}.
\end{equation}
For the collective interaction strength, we find that the following formula determines its value
\begin{align}
    \bar{D}_{\mu}^2 &= \sum_j |\tilde{D}_{\mu,j}|^2 = \frac{|\Omega_{\mu}|^2}{\Delta_{\mu}^2}\sum_j |D_{\mu,j}|^2 = \frac{|\Omega_{\mu}|^2 N }{\Delta_{\mu}^2} \int d\mathbf{r} \tilde{\rho}(\mathbf{r}) |D_{\mu}(\mathbf{r} - \mathbf{r}_c)|^2 = \frac{|\Omega_{\mu}|^2 N}{\Delta_{\mu}^2}I_{\mu}.
\end{align}
The integration $I$ is
\begin{align}
    I_{\mu} &= \int d\mathbf{r} \tilde{\rho}(x, y, z) \frac{C_{3,\mu}^2}{4|\mathbf{r} - \mathbf{r}_s|^6 }(1 - 3\cos^2(\theta))^2 
    \nonumber \\
    & \simeq \frac{8}{(2\pi)^{3 \over 2}\sigma_{\bot}^2 \sigma_z}\int_{-\infty}^{\infty} dz dy  \int_{- \sigma_x}^{ \sigma_x} dx \exp(-\frac{2 z^2}{ \sigma_z^2} - \frac{2 (x^2 + y^2)}{ \sigma_{\bot}^2}) \frac{C_{3,\mu}^2}{(z^2+ y^2+ (x - x_c)^2)^{3}} \left(\frac{ -2 z^2 + y^2 + (x - x_c)^2}{z^2 + y^2 + (x - x_c)^2}\right)^2 \nonumber \\
    & = \frac{2 C_{3,\mu}^2 }{(2\pi)^{3 \over 2}\sigma_{\bot}^5 \sigma_z} \int_{-\infty}^{\infty} d\tilde{z} d\tilde{y}  \int_{-1}^{1} d\tilde{x} \exp(-2 \frac{\sigma_{\bot}^2}{\sigma_z^2}\tilde{z}^2 - 2 \tilde{x}^2 - 2\tilde{y}^2 ) \frac{[-2 \tilde{z}^2 +  \tilde{y}^2 +(\tilde{x} - \tilde{x}_c)^2]^2}{[  \tilde{z}^2 + \tilde{y}^2 + (\tilde{x} - \tilde{x}_c)^2]^{5}} \nonumber \\
    & =\frac{2 C_{3,\mu}^2 }{(2\pi)^{3 \over 2}\sigma_{\bot}^5 \sigma_z} \int_{-\infty}^{\infty} d\tilde{z} d\tilde{y}  \int_{-1}^{1} d\tilde{x} \exp(- \frac{8}{225}\tilde{z}^2 - 2 \tilde{x}^2 - 2\tilde{y}^2 ) \frac{[-2 \tilde{z}^2 +  \tilde{y}^2 +(\tilde{x} - \frac{5}{2 \sqrt{2}})^2]^2}{[  \tilde{z}^2 + \tilde{y}^2 + (\tilde{x} - \frac{5}{2 \sqrt{2}})^2]^{5}} \nonumber \\
      & \simeq \frac{2 C_{3,\mu}^2 }{(2\pi)^{3 \over 2}\sigma_{\bot}^5 \sigma_z} \times 0.136 
\end{align}
If we set $\max(\Omega_{\downarrow}) = 2\pi \times 12.5$ MHz, $\max(\Omega_{\uparrow}) = 2\pi \times 14.7$ MHz, and $N = 500$, we may find 
\begin{equation}
   \max (|\bar{D}_{\uparrow}|) \simeq 2\pi\times 5.64 \ \text{MHz}, \quad \max( |\bar{D}_{\downarrow}|) \simeq 2\pi\times 6.82  \ \text{MHz}.
\end{equation}
The effective Hamiltonian is written as
\begin{align}
    \tilde{H}_{\mu} &= - \sum_j \tilde{\delta}_{\mu,j}  |d_{\mu} \rangle \langle d_{\mu}| \otimes |s_{\mu} \rangle_j \langle s_{\mu} | -( \tilde{D}_{\mu,j} e^{i \mathbf{k}_{\downarrow,0} \cdot \mathbf{r}_j} |d_{\mu} \rangle \langle u_{\mu}| \otimes |g \rangle_j \langle s_{\mu}| + \mathrm{H.c.}).
\end{align}
We expand the wave function as 
\begin{equation}
  |\Psi_{\mu} \rangle = C_{\mu,0} | u_{\mu} \rangle \otimes |G\rangle +  \sum_j C_{\mu,j} e^{-i \mathrm{k}_{\mu,0} \cdot \mathrm{r}_j}|d_{\mu} \rangle \otimes |g \rangle_1 |g \rangle_2 \dots | s_{\mu} \rangle_j \dots |g \rangle_N.  
\end{equation}
The Schr\"{o}dinger equation $i \partial_t |\Psi \rangle = \tilde{H}_1 |\Psi \rangle$ writes as
\begin{align}
    i \partial_t C_{\mu,0} &= \sum_j \tilde{D}_{\mu,j} C_{\mu,j}, \nonumber \\
    i \partial_t C_{\mu,j} &= \tilde{D}_{\mu,j} C_{\mu,0} + (\tilde{\delta}_{\mu,j} - i \Gamma_s/2) C_{\mu,j},
\end{align}
Defining $\tilde{C}_{\mu,j} = e^{\Gamma_s t/2} C_{\mu,j} $, the above equation can be further simplified as 
\begin{align}
    i \partial_t C_{\mu,0} &= \sum_j \tilde{D}_{\mu,j} C_{\mu,j}, \nonumber \\
    i \partial_t \tilde{C}_{\mu,j} &= \tilde{D}_{\mu,j} \tilde{C}_{\mu,0} + \tilde{\delta}_{\mu,j} \tilde{C}_{\mu,j},
\end{align}
The results for simulating this equation are presented in the main text.
We have accounted for the spontaneous decay of the state $|s\rangle$ and set $\Gamma_s \simeq 2\pi \times 5$ kHz.
From the numerical simulation, we get the write efficiency as 
\begin{equation}
    \eta_{\uparrow,w} \simeq 98.93 \%, \quad \eta_{\downarrow,w} \simeq 99.31 \%.
\end{equation}
Then, we estimate the retrieval efficiency $\eta_r$.
Following the expression in Refs. \cite{Ornelas2020Demand,Gorshkov2007Photon, Mei2022Trapped} the value of $\eta_r$ has a relation with the spin wave density as 
\begin{equation}
    \eta_r = \int_0^{1} d\tilde{z} \int_0^{1} d\tilde{z}' \tilde{S}^{*}(1 - \tilde{z})\tilde{S}(1 - \tilde{z}') K_r(\tilde{z}, \tilde{z}'),
\end{equation}
with 
\begin{equation}
    K_r(\tilde{z}, \tilde{z}') = \frac{\text{OD} f(z_s)}{2} \exp{-\frac{\text{OD} f(z_s)}{2}\left[ (1 + z_s (1 - i \Delta) )]\tilde{z} + (1 + z_s (1 + i \Delta) )]\tilde{z}' ] I_0[\text{OD} \sqrt{\tilde{z}\tilde{z}'}f(z_s)\right]},
\end{equation}
and $\tilde{z} = \int_{-\infty}^{z} dz_1 \rho(z_1) $, $z_s = \gamma/|\Omega_c|^2$, $\gamma = (\gamma_s + \gamma_{sg})/\gamma_e$, $f(z_s) = 2/(2 + z_s(1 + \Delta^2))$, $ I_0(z)$ is  $0$th-order modified Bessel function of the first kind, and $\text{OD}$ is the optical deepth.
For OD, it is written as 
\begin{equation}
    \text{OD} = \sigma \int dz  \rho(z) \simeq  \sigma \rho_\text{peak} \sigma_z \sqrt{\frac{\pi}{2}} ,
\end{equation}
with $\sigma$ as the absorption cross section.
As we have supposed, $N = 500$ atoms are Guassian distributing along all direction with $\sigma_x = \sigma_y = 4\ \mu m $, and $\sigma_z = 30\ \mu m $, the peak density is $\rho_\text{peak} = 0.53 \ \mu m^{-3}$. 
If we set the optical cross $ \sigma = 2.9 \times 10^{-13} m^{2}$, we find the value of OD is given by $\text{OD} = 5.79$.
For estimation, we take a one-dimensional approximation for the spin-wave and assume it is given as
$\tilde{S}(\tilde{z}) = 1$ for $\tilde{z} \in [0,1]$ and $\tilde{S}(\tilde{z}) = 0$ elsewhere.
We set the detuning $\Delta = 0$, $\gamma_e = 6.9$ MHz, $\gamma_s = 0$, and $\gamma_{sg} = 0$ MHz.
Combining this value and numerically integrating the equation, we find
\begin{align}
    \eta_r &  \simeq 55.31 \%. 
    \label{eq-retrieval}
\end{align}
Given the write and retrieval efficiency, we get the photon generation probability at the end of the cloud is  
\begin{equation}
   \eta = \eta_{w} \eta_{r} \simeq  54.76 \%. 
\end{equation}
This is comparable to the scheme using a cavity.

Here, we discuss how we can pump the communication atoms from state $|\uparrow \rangle$ to $|u \rangle = |65 P_{3/2}, F = 4, m_F = -4\rangle$ before the atom-ensemble interface is established and from state $|d \rangle = |65 S_{1/2}, F =3,  m_F = -3\rangle$ to the state $|1 \rangle$ after entangling the atom and the ensemble. 
For $^{85}$Rb atoms, the two memory energy levels can be chosen as $|\uparrow \rangle = | 5 S_{1/2}, F = 3,  m_F = -3 \rangle$ and $|\downarrow \rangle = | 5 S_{1/2}, F = 2,  m_F = -2 \rangle $. 
For pumping the communication atom from state $|\uparrow \rangle$ to state $|u \rangle$, we first couple state $|\uparrow \rangle$ to state $|d\rangle$ using a two-photon process as shown in Fig. \ref{supfig2} (a)] 
The single photon detuning is chosen as $\Delta_{mc} = 2\pi \times 100$ MHz.
We can use a $\pi$ pulse to flip the communication atom as $|\uparrow\rangle \rightarrow | d\rangle$. 
Then, we switch off these two lasers and switch on the resonant microwave $\Omega_\text{MW}$ between $|d\rangle$ $|u \rangle$ and have a transition as  $|d\rangle \rightarrow | u\rangle$.
Once the communication atom is pumped to state $|u\rangle$, we switch on the coupling laser within the ensemble and adiabatically convert the ensemble from ground state $|G \rangle$ to spin-wave state $|S\rangle$.
When the spin-wave state preparation process is finished, the communication atom is in state $|d\rangle$.
Then, we utilize the same two-photon process to convert it back to state $|d\rangle$. 
Then, we convert the spin-wave state into a single photon.
The relevant energy levels are shown in Fig. \ref{supfig2} (c).
Here, the intermediate state is $|e\rangle = |5 P_{3/2}, F = 4, m_F = 4 \rangle$.
It can only decay to the ground state $|g \rangle = |5S_{1/2}, F = 3, m_F = 3 \rangle$ and emit a $\sigma^{+}$ photon.

\section{Interaction within the atomic ensemble}

During the analysis, we only take the interaction effects between the communication atom and the atomic ensemble into discussion. 
However, the interaction within the atomic ensemble is neglected. 
Does this interaction affect the generation of entanglement?
To answer this question, let's consider the following Hamiltonian
\begin{align}
    H &= \left(\omega_u | u\rangle \langle u | + \omega_d | d\rangle \langle d |\right) \otimes I_\mathrm{R} + 
 \sum_j I_\mathrm{S} \otimes \left(\omega_s | s\rangle_j \langle s | + \omega_i | i \rangle_j \langle i | + \omega_g | g \rangle_j \langle g | \right) \nonumber \\
    &+ \left( D | u \rangle \langle d | \otimes | i \rangle \langle s | + \Omega(t) I_\text{S} \otimes |i \rangle\langle g| +  \mathrm{H.c.} \right) +  \sum_{j\neq k} I_\mathrm{S} \otimes 
 [V(|\mathbf{r}_j - \mathrm{r}_k|) | s \rangle_j \langle i| \otimes | i \rangle_k \langle s|  + \mathrm{H.c.} ].   
\end{align}
Using the unitary transformation in Eq. \eqref{eq-unitary-transform}, the above Hamiltonian can be transferred into the rotating frame, and the effective Hamiltonian is 
\begin{align}
    H_{\text{eff}} &= \tilde{H}_1 +  \sum_{j\neq k} I_\mathrm{S} \otimes 
 [V(|\mathbf{r}_j - \mathrm{r}_k|) | s \rangle_j \langle i| \otimes | i \rangle_k \langle s|  + \mathrm{H.c.} ]  =  \tilde{H}_1 + \tilde{V}.   
\end{align}
Here $\tilde{H}_1$ is given in Eq. \eqref{eq-effective-three-level}. To characterize the effects of $\tilde{V}$, we verify the states that are scattered from $|u\rangle \otimes | G \rangle $, $|u\rangle \otimes | S \rangle $  by $V$ during the process.
They are
\begin{align}
    V |u \rangle \otimes |G \rangle &= 0, \nonumber \\
    V |d \rangle \otimes |S \rangle &= \sum_{j\neq k} I_\mathrm{S} \otimes 
 [V(|\mathbf{r}_j - \mathrm{r}_k|) | s \rangle_j \langle i| \otimes | i \rangle_k \langle s|  + \mathrm{H.c.} ] \sum_l |d \rangle \otimes |g \rangle_1 |g \rangle_2 \dots |s\rangle_l \dots|g_N \rangle = 0 .
\end{align}
This indicates that although the magnitude of $\mathcal{V}$ is strong as the atoms are confined within a small region, the matrix elements are small.  
It can also be understood from the following fact.
The atoms in the ensembles have dipole-dipole interaction only when one of the atoms is in the Rydberg state $|r_1 \rangle$ and the other atoms are in $|r_2 \rangle$.  
However, during all the dynamics in the main text, the occupation for the intermediate state $|i \rangle$ is small due to the large detuning $\Delta$. 
In total, there is only one single Rydberg excitation in the atomic ensemble. 
Thus, we can neglect the dipole exchange interaction within the ensemble and only consider the interaction between the communication atom and the atomic ensemble. 

The other interaction effect in the atomic ensemble is the blockade effect.
Typically, we utilize $|s \rangle = |65 S_{1/2}, m_j = 1/2 \rangle$ as the target spin-wave state with $C_6 = 360.7$ GHz $\mu m^6$.  
The Rydberg blockade radius is given as 
\begin{equation}
    R_b = (\frac{C_6}{ \Omega_\text{eff}})^{1/6}. 
\end{equation}
If we set $\Omega_\text{eff} = 2\pi \times 1$ MHz, we get $R_b \simeq 6.2\ \mu m$.
For our experiment setup, the ensemble size is much larger than this radius; therefore, the Rydberg blockade effect is not essential in our state preparation process. 

\section{Probability that an atom in the ensemble jumps near the optical tweezer}

Here, we present a discussion regarding the probability that an atom in the atomic ensemble escapes from the trap and interacts with an atom in the optical tweezer. For an atomic ensemble, we trap the atom using the optical dipole trap, which may be approximated as a harmonic trap. If the atoms are in the ground state, we can neglect the interaction between them, yielding the following Hamiltonian
\begin{equation}
    H = \sum_j \frac{1}{2 m}[(p_{x}^{(j)})^2 + (p_{y}^{(j)})^2 + (p_{z}^{(j)})^2] + \frac{m \omega_{\bot} }{2}[(x^{(j)})^2 + y^{(j)})^2] + \frac{m \omega_{z} }{2} (z^{(j)})^2.
\end{equation}
At finite temperature, the density distribution of the atomic ensemble can be calculated from the Boltzmann distribution $f(\mathbf{x}_1, \mathbf{x}_2,\dots, \mathbf{x}_N, \mathbf{p}_1, \mathbf{p}_2,\dots, \mathbf{p}_N ) \propto \exp (- \frac{H}{k_B T})$ as
\begin{align}
    \rho(\mathbf{x}) &= \int \prod_{j=1}^{N}  d\mathbf{x}_j d\mathbf{p}_j f(\mathbf{x}_1, \mathbf{x}_2,\dots, \mathbf{x}_N, \mathbf{p}_1, \mathbf{p}_2,\dots, \mathbf{p}_N ) \sum_k \delta(\mathbf{X} - \mathbf{x}_k) \nonumber \\
    & \propto N \int d\mathbf{x}_1 d\mathbf{p}_1 \delta(\mathbf{X}- \mathbf{x}_1) \exp\{ \frac{1}{2 m k_B T} [(p_{x}^{(1)})^2 + (p_{y}^{(1)})^2 + (p_{z}^{(1)})^2] + \frac{m \omega_{\bot} }{2 k_B T}[(x^{(j)})^2 + y^{(j)})^2] + \frac{m \omega_{z} }{2 k_B T} (z^{(j)})^2\} \nonumber \\
    &\propto N \exp\{ \frac{m \omega_{\bot} }{2 k_B T}(x^2 + y^2) + \frac{m \omega_{z} }{2 k_B T} z^2\},
\end{align}
which is Gaussian. 
We set the density distribution phenomenologically to be the Gaussian function as 
\begin{equation}
    \rho(x, y, z) = \frac{8 N}{(2\pi)^{3/2} \sigma_{\bot}^2 \sigma_z} \exp(- \frac{2 (x^2 +  y^2)}{\sigma_{\bot}^2} - \frac{2 z^2}{ \sigma_z^2}),
\end{equation}
with $\sigma_{z} = 30\ \mu m$ and $\sigma_{\bot} = 4\ \mu m$. For an ensemble with atom number $N = 500$, the peak density is $\rho_\text{peak} = 0.529\ \mu m^{-3}$. 
We place the source atom at the position $(x_c \simeq 7.1\ \mu m,y = 0, z = 0)$ using an optical tweezer. 
The question is to determine how likely it is that the atoms in the ensemble will jump near the optical tweezer, thus making the source atom indistinguishable from the atoms in the ensemble. 
We may think two atoms are indistinguishable if their distance is smaller than $1\ \mu m$. 
Therefore, the number of atoms that may lead to the above event is
\begin{equation}
    P_\text{int} = \int_{x_c-1 \mu m}^{x_c + 1 \mu m} d x  \int_{-1 \mu m}^{1 \mu m} d y \int_{1 \mu m}^{1 \mu m} d z \rho(x, y, z) \simeq N \times 2.4 \times 10^{-5}. 
\end{equation}
For $N = 500$, we get $P_\text{int} \simeq 0.012 \ll 1$. Thus, we can conclude that the atoms in the ensemble are well-trapped and are unlikely to interact with the communication atom in the optical tweezer.

\bibliography{ref}